\documentclass[useAMS,usenatbib,usegraphicx]{mn2e}
\usepackage{color}
\usepackage{lscape}

\newcommand{\kms}{\mbox{km s$^{-1}$}}

\newcommand{\msun}{\mbox{M$_{\odot}$}}

\newcommand{\hii}{\mbox{H~{\sc ii}~}}

\newcommand{\apjs}{\mbox{ApJS}}
\newcommand{\apj}{\mbox{ApJ}}
\newcommand{\mnras}{\mbox{MNRAS}}
\newcommand{\aj}{\mbox{AJ}}
\newcommand{\apjl}{\mbox{ApJL}}
\newcommand{\aap}{\mbox{A\&A}}
\newcommand{\pasj}{\mbox{PASJ}}
\newcommand{\pasp}{\mbox{PASP}}

\title[Young stellar population in Sh2-252]{Young stellar population and ongoing star formation in the \hii complex Sh2-252}

\author[Jose et al.]{Jessy Jose $^{1,2}$\thanks{E-mail: jessy@iiap.res.in}, A.K. Pandey$^{2}$, M.R. Samal$^{3}$, D.K. Ojha$^4$,  K. Ogura$^5$,  J.S. Kim$^6$, 
\newauthor
N. Kobayashi$^{7}$, A. Goyal$^{8}$, N. Chauhan$^{9}$ and C. Eswaraiah$^2$\\
$^1$ Indian Institute of Astrophysics, Koramangala, Bangalore, 560 034, India\\
$^2$ Aryabhatta Research Institute of observational sciencES (ARIES), Manora Peak, Naini Tal, 263129, India\\
$^3$ Aix Marseille Universit\'e, CNRS, LAM (Laboratoire d'Astrophysique de Marseille), UMR7326, 13388 Marseille, France\\
$^4$ Tata Institute of Fundamental Research, Mumbai (Bombay), 400 005, India\\
$^5$ Kokugakuin  University, Higashi, Shibuya-ku, Tokyo, 150-8440, Japan\\
$^6$ Steward Observatory, 933 North Cherry Avenue, Tucson, Arizona, 85721-0065, USA\\
$^7$ Kiso Observatory, School of Science, University of Tokyo, Mitake, Kiso-machi, Kiso-gun, Nagano-ken 397-0101, Japan\\
$^8$ National Centre for Radio Astrophysics/TIFR, Pune University Campus, 411 007, Pune, India\\
$^9$ Institute of Astronomy, National Central University, Chung-Li, 32054, Taiwan\\
}
\begin{document}

\date{}

\pubyear{2013}

\maketitle

\label{firstpage}

\begin{abstract}
In this paper, an extensive survey of the star forming complex 
Sh2-252 has been undertaken with an aim to explore its hidden young stellar population
as well as to understand the structure and star formation history  for the first time. 
This complex is composed of five prominent embedded clusters associated with the sub-regions A, C, E,
NGC 2175s and Teu 136. We used 2MASS-NIR and {\it Spitzer}-IRAC, MIPS photometry  to identify and classify the 
young stellar objects (YSOs) by their infra-red (IR) excess emission. Using the  IR colour-colour criteria, we identified 577
YSOs,  of which, 163 are Class I, 400 are Class II and 14 are transition disk YSOs, 
suggesting a moderately rich number of YSOs in this complex. Spatial distribution of the 
candidate YSOs shows that they  are mostly clustered around the sub-regions in the western 
half of the complex, suggesting enhanced star formation activity towards its west. 
Using the spectral energy distribution (SED) and optical colour-magnitude diagram (CMD)  
based age analyses, we derived probable evolutionary status 
of the sub-regions of Sh2-252. Our analysis shows that the region A is the youngest ($\sim$ 0.5 Myr), 
the regions B, C and E are of similar  evolutionary stage ($\sim$ 1-2 Myr) and the  clusters 
NGC 2175s and Teu 136  are slightly evolved ($\sim$ 2-3 Myr). Morphology of the region in 
the 1.1 mm map shows a semi-circular shaped molecular shell composed of several clumps and 
YSOs  bordering the western ionization front (IF) of Sh2-252. Our analyses suggest that next 
generation star formation is currently under way along this border and that possibly 
fragmentation of the matter collected  during the expansion of the \hii region as one of the major
processes responsible for such stars. 
We observed the densest concentration of YSOs (mostly Class I, $\sim$ 0.5 Myr) 
at the western outskirts of the complex, within a  molecular clump associated with  water 
and methanol masers and we suggest that it is indeed a site of cluster formation at a very 
early evolutionary stage, sandwiched between the two relatively evolved C\hii regions A and B. 

\end{abstract}


\section{Introduction}
\label{intro}

In order to understand the interstellar processes which govern  star formation, studies of the large scale 
properties of various star forming regions at different environments are very essential. Census of young stellar 
objects (YSOs) born in a cloud, their mass, age and spatial distribution, together with  mass of the star 
forming cloud are some of the key data used for the assessment  of the time scale and  star formation 
history of a given region.  However, identifying the member stars of a given young association is quite difficult. 
It is well known that  YSOs show excess emission  in infra-red (IR) above main sequence (MS)  photosphere due to 
thermal emission from their circumstellar material. 
Thus, YSOs can be identified by looking for IR excess emission.  YSOs are often  categorized  into 
Class 0, I, II or III evolutionary stages \citep{lada1984}. Class 0 objects are deeply embedded protostars that are still 
experiencing cloud collapse. They are extremely faint at wavelengths shorter  
than 10 $\mu$m and  have a significant sub-millimeter luminosity.  A Class I YSO is also an object whose 
emission is dominated by a dense infalling spherical envelope,  however these objects are bright in the IR.
A Class II YSO is characterized by the presence of an optically thick, primordial circumstellar disk, which 
dominates the star's emission. When the  circumstellar disk material  becomes optically thin, the star 
is classified as a Class III star. Hence, IR survey of a star forming region can be
used as a powerful tool to distinguish the stars with IR excesses from stars without such excesses.  
The unprecedented sensitivity and mapping capabilities  of the {\it Spitzer} 
Space Telescope provides an excellent platform to survey the star forming regions in the mid-IR and
to identify the  IR  excess sources in the region.  Furthermore,  {\it Spitzer}  data can also be used to classify  
YSOs at  different evolutionary stages,  which helps us probing the  recent star formation activity of a given region (e.g., \citealt{smith2010}). 

 Massive stars have a profound effect on their natal environment creating wind-blown shells, cavities and HII regions. 
The immense amount of energy released through their stellar winds and  radiation disperse and destroy the remaining molecular
gas and likely inhibit further star formation.  However, it has also been argued that in some circumstances, 
the energy input by these massive stars can promote and induce subsequent low mass star formation in the surrounding molecular gas before
it disperses (e.g., Koenig et al. 2012).  Therefore, regions that contain massive stars provide excellent laboratories for studying both
high and low mass star formation. Identification and characterization of the YSOs in star forming    complexes hosting massive stars are 
essential steps to examine the    physical processes that govern new  generation star formation  in such complexes.

\begin{figure*}
\centering
\includegraphics[scale = 0.8, trim =0 20 20 230, clip]{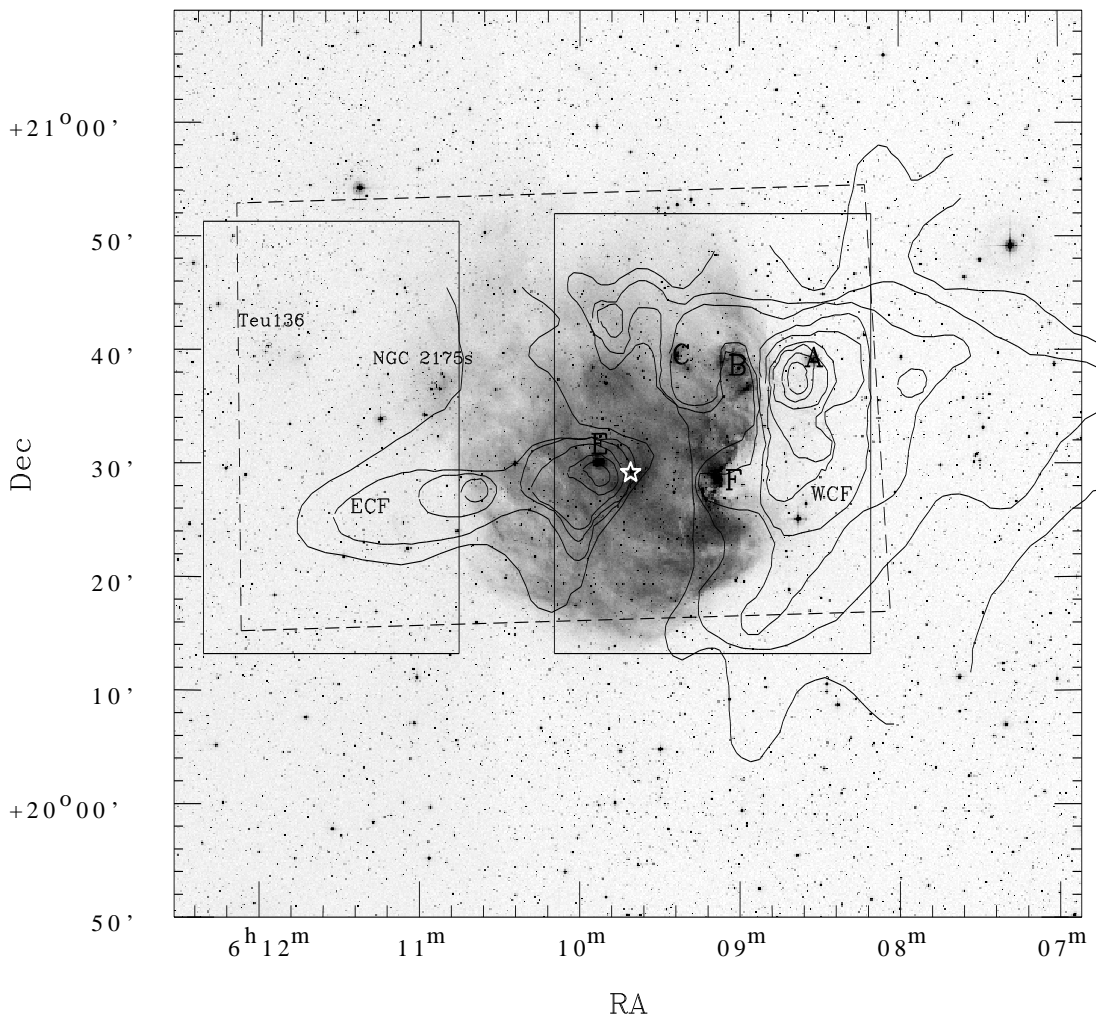}
\caption {The DSS2-R band image of the region around Sh2-252  for an area of $\sim$ 1.3  $\times$ 1.3  deg$^2$. 
The locations of the thermal radio sources A, B, C, E and F identified by \citet{felli1977}, 
and the small clusters NGC 2175s and Teu 136 are marked in the figure. The ionizing source (HD 42088) of the \hii region is 
marked by using a star symbol. The contours are the $^{12}$CO Rayleigh-Jeans brightness temperature 
map taken from \citet{lada1979}. The contours are shown with an interval of 3 K, where the lowest contour 
corresponds to the temperature of 6 K.  The region shown using the  dashed box is covered by the {\it Spitzer}-IRAC 
observations and  the regions shown by solid boxes represent the area covered for the {\it Spitzer}-MIPS-24 $\mu$m observations.  
The details of the WCF and ECF are given in the text. }
%The abscissa and the ordinate are for the J2000 epoch. }
\label{area}
\end{figure*}

Here we present the results of our multi-wavelength analyses of the  star forming complex Sh2-252 (\citealt{sharpless1959};
$\alpha_{2000}$ = $06^{h}09^{m}39^{s}$; $\delta_{2000}$ = $+20^{\circ}29^{\prime}12^{\prime\prime}$; l=190$^\circ$.04; 
b=+0$^\circ$.48). Sh2-252 is an evolved \hii region powered by the central massive ionizing source of O6.5V type, HD 42088, 
which is a member of the Gemini OB1 association. The DSS2-R band image of the region around Sh2-252 for an area 
of $\sim$ 1.3  $\times$ 1.3  deg$^2$ is shown in Fig. \ref{area}. Radio observations at 5 GHz  detected six 
extended sources (Sh2-252 A to F) at different spatial locations  towards the region,  of which, four of them 
(A, B, C and E) are classified as compact \hii (C\hii) regions (\citealt{felli1977}, \citealt{lada1979}). The 
stellar contents of this region were not so well explored until recently. \citet{jose2012} (hereafter Paper 1) 
unraveled its stellar contents using deep optical, shallow near-infrared (NIR; 2MASS), $H\alpha$ survey  along 
with  spectroscopic observations. They estimated the distance (2.4 kpc), reddening and identified 12 OB stars of 
spectral type earlier than B8, 61 $H\alpha$ emission line sources  within the nebula and also 5 
prominent embedded clusters associated with the sub-regions A, C, E, NGC 2175s and Teu 136. The age and age spread, 
initial mass function (IMF), $K$-band luminosity function (KLF) etc. of the sub-regions  have also been discussed in Paper 1. 
The main sub-regions of Sh2-252 such as A, B, C, E, F,  NGC 2175s and Teu 136 are marked in Fig. \ref{area}. 
Sh2-252 is bounded by an ionization front (IF) on the west side, while it is density bounded on the east side 
\citep{felli1977}. The detailed $^{12}$CO and $^{13}$CO maps of the region by \citet{lada1979} showed  that the cloud 
complex is separated  into two fragments by a long rift of little CO emission and named them as western cloud 
fragment (WCF) and eastern  cloud fragment (ECF),  respectively, and are marked in Fig. \ref{area}.  Most of the
mass of WCF is concentrated in a narrow ridge which borders the IF of Sh2-252  at its west (\citealt{lada1979}; \citealt{kompe1989}). 
The similarity in the kinematics of these two cloud fragments  shows that they belong to the same cloud complex and 
ECF probably lies at the back edge of the \hii region, though it is projected very close to the
ionizing source (\citealt{lada1979}; \citealt{fountain1983}). The most intense CO peak of the WCF 
is located very close to   region A, with  water and methanol maser emissions (\citealt{lada1979}; 
\citealt{kompe1989}; \citealt{szymczak2000}) within its  proximity,  suggesting  recent  star formation
activity towards this region (\citealt{lada1979}; \citealt{lada1981}; \citealt{kompe1989}).  
An extensive overview of the previous studies of Sh2-252  can  be found in Paper 1.

This work is the continuation of Paper 1,  where we have  focused  mainly on the  characteristics of the 
YSOs in the Sh2-252 complex for the first time  and to constrain the star formation history of the complex. We used  
the optical, 2MASS-$JHK$, {\it Spitzer}- IRAC and MIPS data sets to identify and classify the YSOs of Sh2-252 region. 
An attempt has also been made to characterize the physical properties of the YSOs based on the spectral energy 
distribution (SED) analyses.  Section 2 
describes various data sets used for the present study.  Analyses and results including the identification 
and classification of YSOs, their various properties such as ages, 
masses, SEDs,  spatial distribution   etc. have been  discussed in Section 3. Section 4 deals with   discussion on the
morphology and star formation scenario of the complex and finally  summary of the paper is  presented in Section 5.

%%%%%%%%%%%%%%%%%%%%%%%%%%%%%%%%%%%%%%%%%%%%%%%%%%%%%%%%%%%%%%%%%%%%%%%%%%

%%%%%%%%%%%%%%%%%%%%%%%%%%%%%%%%%%%%%%%%%%%%%%%%%%%%%%%%%%%%%%%%%%%%%%%%

\section{Data sets}

\subsection{{\it Spitzer}-IRAC observations}

Two sets of IRAC observations in 3.6, 4.5, 5.8 and 8.0 $\mu$m bands (channels 1, 2, 3 and 4)
were available in  the {\it Spitzer} space  observatory archive
program (Program IDs: 201; 20506, PIs: G. Fazio; J. Hester) which were taken towards WCF and ECF
on 2004 October 28 and  2006 March 26, respectively.  The IRAC data
were taken  in the High Dynamic Range  mode  with three dithers
per map position and with integration time of 0.4 sec and 10.4 sec per
dither. We downloaded the basic calibrated data  images 
(versions S18.7.0 and S18.5.0) from the {\it Spitzer}
archive\footnote {http://archive.spitzer.caltech.edu/} and the raw
data were processed and calibrated with the IRAC pipeline. The final
mosaics were created using the MOPEX  pipeline (version 18.0.1) with
an image scale of 1$^{\prime\prime}$.2 per pixel. The total area
covered by these two  observations is marked in Fig. \ref{area} by using
a dashed box.  

We performed the point response function (PRF) fitting method in multiframe mode  using the tool
APEX developed by {\it Spitzer} Science Center  on all the  {\it   Spitzer}-IRAC images  
to extract the magnitudes. The standard PRFmap table\footnote
{http://ssc.spitzer.caltech.edu/irac/calibrationfiles/psfprf/prfmap.tbl}
provided on the {\it Spitzer} website   were used to fit variable PRFs
across the image. Point sources with peak values more than 5$\sigma$ above
the background were considered candidate detection. Many sources detected in the nebulosity were appeared to
be spurious.  The spurious sources were  identified visually and deleted  from the
automated detection list. Also, some of the sources were not detected
by APEX automatically.  We  identified  these  sources visually and
added  the coordinates in the user-list mode in APEX to extract the
magnitudes and thus we made sure that photometry of every genuine
source is derived.  We have adopted the zero-points for conversion
between  flux densities and magnitudes to be 280.9, 179.7, 115.0 and
64.1  Jy in the 3.6, 4.5, 5.8 and  8.0 $\mu$m bands, respectively,
following the IRAC Data Handbook. There were signs of saturation of
bright sources in the long integrated images. The photometric data derived
from the short images were replaced for the saturated sources. The two
sets of IRAC observations had substantial overlapping between the
regions and we used the  average magnitudes of  those
common sources detected in both the  observations. Channels 3 and 4
detected far fewer sources than channels 1 and 2. This deficit exists
because channels 1 and 2 are more sensitive than channels 3 and 4 and
are also less affected by the bright diffuse emission  that dominates
the channel 3 and 4 images.  The underlying typical stellar photosphere is
also intrinsically fainter at channels 3 and 4 than at channels 1 and 2,
which further hinders the ability to detect sources in the longer wavelength
IRAC filters. We finally obtained extraction lists of
24877, 20591, 5083 and 3263 sources  in the IRAC 3.6, 4.5, 5.8, and
8.0 $\mu$m bands, respectively, within the area shown in Fig. \ref{area}. 
The IRAC data of the four band passes were merged,  by  matching the coordinates 
using a radial matching tolerance of 1$^{\prime\prime}$.2. 
To ensure good photometric accuracy,  we restricted our catalog with sources 
having uncertainty $\le$ 0.2 mag in all the bands. Thus our final IRAC catalog  
contains photometry of 31251  sources which were detected in one or
more IRAC bands. Out of these, only 1988  sources were detected
in all the IRAC  bands.  

\subsection{{\it Spitzer}-MIPS 24 $\mu$m observations}

The areas towards WCF and ECF were observed at 24 $\mu$m  using the Multi-band
Imaging Photometer for {\it Spitzer} (MIPS)    on 2006 October 13  and 2008 October 25
(Program IDs: 20726; 50758, PIs: J. Hester; M. M. Drosback). We downloaded the BCD images 
(version S18.12.0) from the {\it Spitzer} archive and the final mosaics were created using the
MOPEX  pipeline (version 18.0.1) with an image scale of
2$^{\prime\prime}$.45 per pixel. The total area covered by these
observations is marked in Fig. \ref{area}. The sources were identified visually and their coordinates  
were given  in the user-list mode in APEX. To
extract the flux, we performed the PRF fitting method in the single frame
mode.  The zero-point value of 7.14 Jy from the MIPS Data Handbook
has been used  to  convert flux densities  to magnitudes. The final
catalog contains  the 24 $\mu$m photometry of 646 sources having
uncertainty $\le$ 0.2 mag. We cross-correlated our MIPS-24 $\mu$m
catalog with the source catalog of IRAC, using a
2$^{\prime\prime}$.5 matching  radius. We obtained 24 $\mu$m counterparts
for 428 sources with IRAC data in all the bands. The 24 $\mu$m photometry was mainly used to
classify the YSOs and to analyze their SEDs (see Section \ref{sed}).

\subsection{Near-infrared data from 2MASS}

NIR $JHK$ data for  point sources within a radius of 30 arcmin
around HD 42088  have been obtained from  2MASS point source catalog (PSC) 
\citep{cutri2003}.  To improve the photometric accuracy, we used the data with 
photometric quality flag (ph$\_$qual = AAA) which gives a S/N $\ge$ 10 and a  
photometric uncertainty $ <$ 0.1 mag. This selection criterion ensures best 
quality detection in terms of photometry and astrometry as given on the  2MASS
website\footnote {http://www.ipac.caltech.edu/2mass/releases/allsky/doc/}. We used 2MASS
data  along with {\it Spitzer}-IRAC data  to identify and classify the 
candidate YSOs within Sh2-252 (see Section \ref{yso_2mass}). The NIR and
IRAC sources were  bandmerged  within $1^{\prime\prime}$.2 radial
matching tolerance. If more than one 2MASS sources satisfied this
requirement, then the nearest one was selected. Thus we obtained a total of  9097
2MASS sources having IRAC  counterpart at least in one band.

\section{Results} 

\subsection{General morphology of the region}
\label{morphology}

\begin{landscape}
\begin{figure} 
\centering

\includegraphics[scale = 1.5, trim = 0 100 0 0, angle=0,   clip]{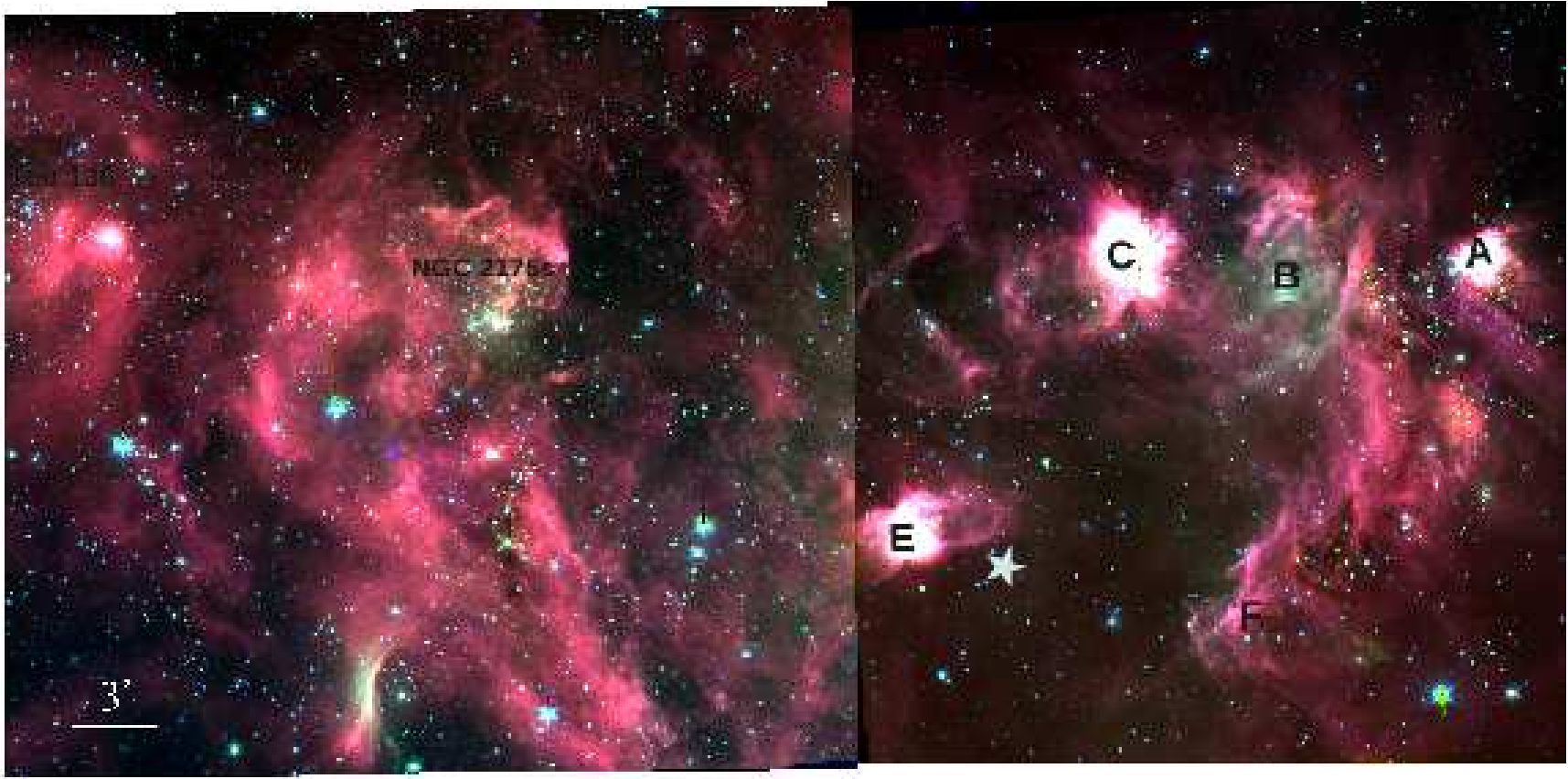}
\caption{IRAC three colour image of Sh2-252 (blue: 3.6 $\mu$m; green: 4.5 $\mu$m; red: 8.0 $\mu$m; see the electronic version for the colour image).
North is up and east is to the left. The important 
sub-regions are marked in the figure and the white star symbol represents the location of HD 42088, the ionizing 
source of Sh2-252.}

\label{color-irac}
\end{figure}
\end{landscape}

  A colour composite image using the 3.6, 4.5 and 8.0 $\mu$m bands of {\it Spitzer}-IRAC
  is shown  in Fig. \ref{color-irac}.  The important sub-regions and the ionizing source (HD 42088) of
Sh2-252 are marked in the figure.  Different IRAC bands have   contribution from various  extended emission 
features.  For example,   3.6 $\mu$m band is  dominated by diffuse emission from  the 3.3
  $\mu$m PAH feature.  The 4.5 $\mu$m band includes the $Br{\alpha}$  line plus CO band head emission and 
$H_2$ lines, and 5.8 and 8.0   $\mu$m  bands are dominated by strong PAH features,  although a few
  sources show extended thermal continuum emissions from warm   circumstellar  dust in these bands \citep{smith2010}.
  PAHs are believed to be destroyed in the ionized gas and thought to be excited in the
  photon dominated region (PDR)   that is situated at the interface of \hii region and molecular cloud by the absorption of
 far-UV photons leaking from the \hii region \citep{pomares2009}. Therefore PAHs are good tracers of the warm PDR that 
 surrounds the \hii region. The most   striking feature in  Fig. \ref{color-irac} is the magenta coloured
  emission that appears almost semi-circular in shape towards the west of Sh2-252. The PAHs within the PDR  
re-emit their energy in   the 3.6, 5.8 and 8.0 $\mu$m bands, and this can be seen as  magenta
  coloured emission  in  the colour composite image given   in Fig. \ref{color-irac}.  Altogether, 
this image shows a bubble-like emission feature around the   ionizing source HD 42088, likely to be  created 
due to the expansion of the   \hii region to its surroundings. The bright-rim feature discussed in Section \ref{intro}
  (i.e., region F; see also Fig. 2 of  Paper 1 for a close-up view) coincides well with the PAH emission feature
  seen at the southwest of Fig. \ref{color-irac},  with a sharp edge   pointing towards HD 42088. This sharp edge  
indicates the IF,   that coincides with the  molecular ridge seen in the $^{12}$CO map shown in Fig. \ref{area} 
  as well as the extinction map shown in Fig. \ref{extmap} (see Section \ref{kextmap}). This correlation indicates the interface  
  of the ionized and molecular gas. The correlation of the half-ring structure of PAH emission towards the west of 
Sh2-252 with the $^{12}$CO emission and extinction maps indicates that the ionized gas is separated from the molecular 
cloud by the  neutral gas   with PAH emission.   The absence of 8 $\mu$m emission in the interior of the \hii region can
  be interpreted as the destruction of PAH molecules by the intense UV radiation of the ionizing source. There is also 
less intense diffuse emission towards the east of Sh2-252. However, the ionized gas is  bounded more extensively in the 
western  direction than in the  eastern direction. It is to be noted that the location of NGC 2175s is devoid of strong 
PAH emission features. Interestingly,  another bright red emission knot is seen further towards east of NGC 2175s with 
a bright red point   source inside it.  A small clustering is evident further east to this point source which coincides 
with the sub-region Teu 136 discussed in Paper 1. There are not many studies on this group of stars in the literature. 
However, based on our spectral analysis in Paper 1, we concluded that  this group is a part of the main Sh2-252 complex itself. 

%%%%%%%%%%%%%%%%%%%%%%%%%%%%%%%%%%%%%%%%%%%%%%%%%%%%%%%%%%%%%%%%%%%%%%%%%%
%%%%%%%%%%%%%%%%%%%%%%%%%%%%%%%%%%%%%%%%%%%%%%%%%%%%%%%%%%%%%%%%%%%%%%%%  

\subsection{$K$-band extinction map from H and K data} 
\label{kextmap}
%%%%%%%%%%%%%%%%%%%%%%%%%%%%%%%%%%%%%%%%%%%%%%%%%%%%%%%%%%%%%%%%%%%%%%%%%%
\begin{figure*}
\centering
\includegraphics[scale = 0.6, trim = 0 0 0 0, clip]{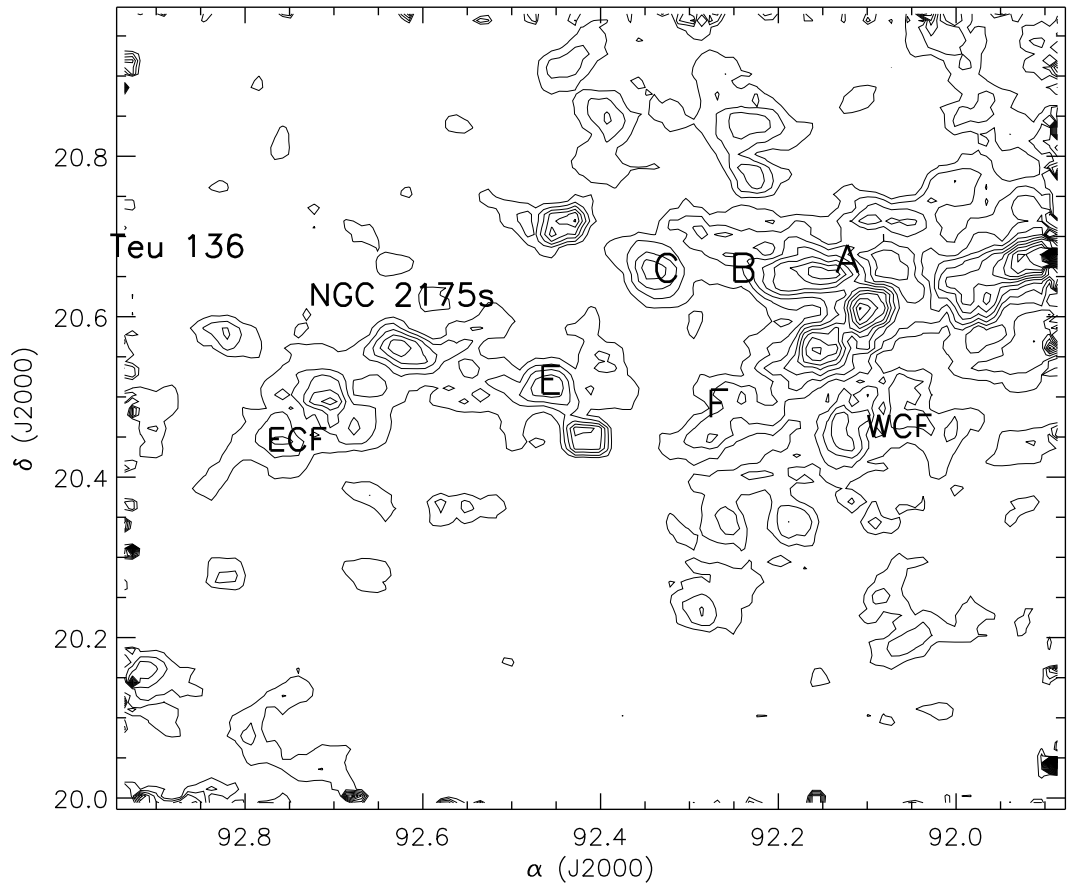}
\caption{$K$-band extinction map generated from the $(H-K)$ colours of 2MASS data. The contours begin at $A_K$ = 
0.55 mag and increase by 0.15 mag up to 1.6 mag. The locations of the important sub-regions of Sh2-252 are marked
in the figure. The abscissa and the ordinate are for the J2000 epoch.
}
\label{extmap}
\end{figure*}
%%%%%%%%%%%%%%%%%%%%%%%%%%%%%%%%%%%%%%%%%%%%%%%%%%%%%%%%%%%%%%%%%%%%%%%%

The indication of variable extinction in the Sh2-252 region has been discussed in Paper 1
based on the optical photometric analyses of the bright stars as well as from the distribution 
of candidate YSOs on the $(J-H)/(H-K)$ colour-colour (C-C) diagram.  In order to quantify
the amount of extinction within each sub-region of Sh2-252 and to characterize the structure of 
molecular clouds, we derived the $A_K$ extinction  map using the $(H-K)$ colours of the stars in the 
2MASS catalog. To improve the quality of the map, the candidate YSOs and probable contamination 
sources (see Section \ref{yso}) were excluded from the list.  We used the grid method to 
determine the mean value of $A_K$. Briefly, we divided the Sh2-252 region into many
grids of size $30^{\prime\prime} \times 30^{\prime\prime}$, and the median value of  $(H-K)$ 
colours of all the sources  within each grid was measured. The sources deviating  above 3$\sigma$  
were excluded to calculate the final median colour of each grid. We used the reddening law by \citet{flaherty2007} to
convert  $(H-K)$  colour in to $A_K$,  using the relation  $A_K$ = 1.82 $\times$
$(H-K)_{obs}-(H-K)_{int}$, where, $(H-K)_{int}$ = 0.2 is assumed as an
average intrinsic colour of stars in young clusters (see \citealt{allen2008};
\citealt{gutermuth2009}).  Since this complex is situated at a
distance of 2.4 kpc (Paper 1), the contribution from the foreground field stars
would have affected the present extinction measurements. An average
foreground extinction of  $A_V$= 1.5 mag has been estimated towards this region in Paper 1. 
Therefore, to eliminate the  foreground  contribution in the extinction measurement,
we used only those stars with $A_K$ $>$ 0.15 mag to generate the
extinction map.  There were an average of five 
sources in each grid to calculate the median colour. The final extinction map, having an angular
resolution of $30^{\prime\prime}$, which is sensitive down to $A_V$ = 16
mag  is shown in Fig. \ref{extmap}. However, it has to be kept in mind
that this map is limited by the sensitivity of the 2MASS survey. As a
result, the $A_K$ value derived may be underestimated  due to the
small number of background stars detected in the  heavily extincted area of the complex.  

It is interesting to note that the extinction map derived from the $(H-K)$ colours 
shown in Fig. \ref{extmap}  resembles the general distribution of the molecular cloud
as outlined by the $^{12}$CO emission map shown in Fig. \ref{area}. 
It traces the  presence of molecular cloud towards the sub-regions A, B, C and nicely follows
the bright rim feature  visible at the west of Sh2-252 (i.e., region F, see Fig. 2 of Paper 1) similar to the
ridge of molecular cloud seen in the $^{12}$CO emission map by \citet{lada1979}. 
Similarly, the separation between the two cloud fragments WCF and ECF (see Section \ref{intro})
by the long rift of less extinction is also clearly seen in Fig. \ref{extmap}.
The western cloud portion is more extincted as compared to the eastern one.
However, isolated extinction  complexes are seen in the eastern side as
well. It has to be noted that the locations  of the clusters NGC 2175s  and Teu 136 are
devoid of heavy reddening. The same trend has already been noticed in Paper 1.  The average extinction 
$A_V$  of the four optically bright members of the cluster NGC 2175s (see Table 3 of Paper 1) is
$\sim$ 2.2 mag,  which is significantly lower than the extinction towards the western part of the
complex. This could be explained by the fact that the strong   stellar  winds
from the four early type sources in NGC 2175s would have removed the
dust and gas from its neighborhood or the region could be more evolved than the other parts of the complex. 

\subsection{Identification of YSOs based on near- and mid-infrared colours}
\label{yso}

In the ensuing sections, we identify and classify the YSOs within Sh2-252 mainly into Class I and Class II 
categories using their IR colours. We do not classify diskless Class III YSOs  because they are indistinguishable 
from field stars based on their IR colours and hence  we  cannot reliably identify  them using the current data set.
The main limitation of the identification of YSOs based on their IR colours is the contamination from 
various non-stellar  sources in IRAC detection such as,
polycyclic aromatic hydrocarbon (PAH) emitting  galaxies, broad-line
active galactic nuclei (AGNs), unresolved knots of shock emission, PAH-emission contaminated
apertures etc., which have colours similar to that of YSOs and may lead  to wrong identification of YSOs. 

\citet{megeath2004} and \citet{allen2004} introduced the
classification criteria for YSOs using the ($[3.6]-[4.5]$, $[5.8]-[8.0]$)
C-C diagram.  These colour criteria since then  have been
modified by several authors (e.g.,  \citealt{harvey2006, harvey2007},
\citealt{gutermuth2008, gutermuth2009}, \citealt{evans2009} and the references therein).
We adopted  the empirical three-phase scheme introduced by \citet{gutermuth2009} to identify the YSOs with IR
excess in this complex.  Briefly, in phase 1, after excluding the non-stellar contamination (see Section \ref{yso_irac}),
 all four IRAC bands are used to separate the sources with IR excesses due to disks and  envelopes (Class II and
Class I, respectively) from pure photospheres (Class III/field
stars) based on [3.6] - [5.8] and [4.5] - [8.0] colours. 
Phase 2 is applied to those sources which lack IRAC detection at [5.8] or [8.0] $\mu$m, but which have high-quality 2MASS
detection ($\sigma <$ 0.1 mag) in at least $H$ and $K$ bands. Sources
are then classified as  Class II or Class I objects based on their  de-reddened
($([K]-[3.6])_0$ and  $([3.6]-[4.5])_0$) colours \citep{allen2008}.
In addition, there are a few   objects with NIR+IRAC
colours of Class IIIs, but which exhibit a 24 $\mu$m excess.  These objects
are included in a new class of YSOs i.e., `transition disk' sources.
Phase 3  re-examines all sources including contaminants and identifies
Class I objects and transition disk objects on the basis of
their MIPS  24 $\mu$m photometry.

\subsubsection{Selection of YSOs using IRAC data}
\label {yso_irac}

%%%%%%%%%%%%%%%%%%%%%%%%%%%%%%%%%%%%%%%%%%%%%%%%%%%%%%%%%%%%%%%%%%%%%%%%%%
\begin{figure*} 
\centering
\includegraphics[scale = 0.9, trim = 0 40 0 220, clip]{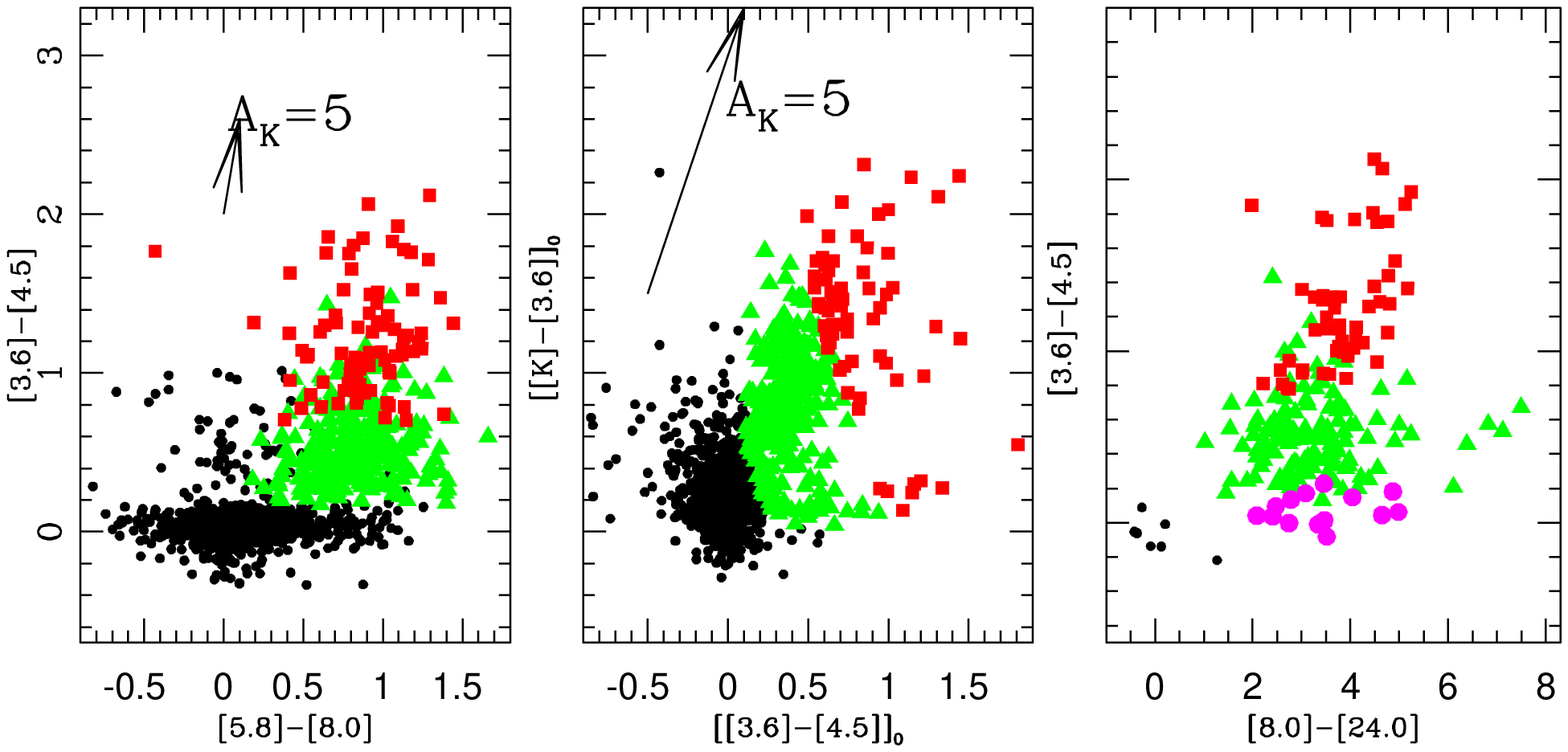}

\caption{{\it left}: IRAC C-C diagram of all the uncontaminated sources identified  
within the region   shown by a dashed box in Fig. \ref {area}. The YSOs classified as Class I and Class II, 
based on the colour criteria  by \citet{gutermuth2009}, are shown by using red squares and green triangles, respectively. 
The reddening vector for $A_{K}$ = 5 mag is plotted by using the reddening law from \citet{flaherty2007}. 
{\it middle}:  The dereddened $([K]-[3.6])_0$ vs $([3.6]-[4.5])_0$ C-C diagram for all the IRAC sources having 2MASS 
counterparts after removing the contaminants. The sources in red  and green  are the candidate Class I 
and Class II YSOs, respectively. {\it right}: [3.6]-[4.5] vs [8.0]-[24] C-C diagram for all the IRAC sources having 24 $\mu$m
photometry after removing the contaminants. The Class I, Class II and transition disk YSOs are shown by using  red squares, green  
triangles and magenta circles, respectively. }

\label{irac-cc}
\end{figure*}
%%%%%%%%%%%%%%%%%%%%%%%%%%%%%%%%%%%%%%%%%%%%%%%%%%%%%%%%%%%%%%%%%%%%%%%%  

 In order to identify the YSOs by their IRAC colours, we used only those sources detected in all
the IRAC bands with photometric uncertainties less than 0.2 magnitude.
  Various non-stellar contaminating sources mentioned above are eliminated by
using  several criteria described in detail by \citet{gutermuth2009}. Out of the  1988  sources 
detected in all the IRAC channels,  444  sources are found to be contaminants.  The remaining 
sample of 1544 sources was used to identify the   Class I  and  Class II YSOs  associated with the region using the IRAC
 colour criteria given by \citet{gutermuth2009}.  The above criteria yields 340 YSOs with IR
excess emission,  with 86 and 254 sources having colours consistent with Class I
and Class II types, respectively. The rest could be Class III/field stars, which  we 
are unable to classify reliably based on their IRAC colours.
  However, it is to be noted that in a few cases, highly
reddened Class II sources could mimic  the colours of  Class I source
\citep{gutermuth2009}.  In Fig. \ref{irac-cc} (left panel), we have
shown the $[3.6]-[4.5]$ vs $[5.8]-[8.0]$ C-C diagram for all the
uncontaminated IRAC sources, where the Class I and Class II sources
are shown as red squares and green triangles, respectively.  Here we would like to 
point out that the present YSO selection based on the IRAC- four colour detection
is  not complete, as many sources falling in the bright nebulous regions
were not detected in the 5.8 and 8.0 $\mu$m  bands.

\subsubsection{Additional YSOs from $H, K, 3.6, 4.5$   and $24$ $\mu$m data}
\label{yso_2mass}

The detection of Class I and Class II YSOs in all the four IRAC bands is limited mainly by the 
lower sensitivity and enhanced  nebulosity of the [5.6] and [8.0] $\mu$m bands. 
In order to account for the  YSOs  missing in these bands, we used   $H$, $K$, [3.6] and [4.5] $\mu$m  data sets 
and the additional YSOs in the complex are identified.
\citet{gutermuth2009} used the dereddened colour criteria to identify such YSOs.  We used
the extinction map (see Fig. \ref{extmap}; Section \ref{kextmap}) to deredden the
data.  To minimize  the inclusion of extra-galactic contaminants,
we applied a simple brightness limit in the dereddened [3.6] $\mu$m
photometry  i.e.,  all the Class I YSOs must have $[3.6]_0 \le$ 15 mag
and all the Class II YSOs must have $[3.6]_0 \le$ 14.5 mag \citep{gutermuth2009}.
Thus, after removing the contaminants and using the \citet{gutermuth2009}
colour criteria, we identified a total of 367 candidate   YSOs (67
Class I; 300 Class II)  with IR excess emission,  of which,  211
(52 Class I and 159 Class II) are found to be common with the YSOs
identified using the IRAC four-colour criteria  (see
Section \ref{yso_irac}). Finally, we obtained 15 and 141 additional YSOs   having
colours consistent with Class I and  Class II, respectively using the $H, K, [3.6]$ and $[4.5]$ 
colour combination. In Fig. \ref{irac-cc} (middle
panel), we have shown the  $([K]-[3.6])_0$ vs $([3.6]-[4.5])_0$ C-C
diagram for all the uncontaminated sources, where the Class I and
Class II sources  are shown in red squares and green triangles,
respectively.

In order to identify the transition disk sources, we re-examined the entire catalog of  sources 
having 24 $\mu$m counterparts and  with uncertainty less than 0.2 mag. 
Those sources that were considered photospheric in the previous classification but have
significant excess emission at  24 $\mu$m (i.e., [5.8]-[24] $>$ 2.5 mag or [4.5]-[24] $>$ 2.5 mag)
are typically known as transition disks. They are the  Class II sources with significant dust 
clearing within their inner disks \citep{gutermuth2009} and we identified 14 such sources within the complex. 
Also, any source which lacks detection in some IRAC bands yet has bright  24 $\mu$m photometry 
are likely to be deeply embedded protostars. We identified 67 such sources and those are included  
in the Class I category in the following analysis.  Five  sources which lacked  strong 24 $\mu$m 
detection but had been classified as  Class I YSOs in the previous step (Section \ref{yso_irac}) are
reclassified as heavily reddened Class II YSOs.

Thus after  reclassifying the sources using the  24 $\mu$m photometry, we have finally identified
577 YSOs with IR excess emission in the Sh2-252 complex. Of these, 
163 are consistent with  Class I, 400 are consistent with Class II and 14 are consistent with
transition disk YSOs. These objects will be used in the following analysis. Thus the Sh2-252 complex seems to have 
moderately rich number of YSOs. In the right panel of  Fig. \ref{irac-cc} we have shown the 
[3.6]-[4.5] vs [8.0]-[24] C-C diagram of all the uncontaminated 
IRAC sources having counterparts in 24 $\mu$m along with the candidate
Class I, Class II and transition disk  sources marked in red squares, green  triangles and magenta circles,  respectively.

\subsection{Completeness of the census of YSOs}

\begin{figure*}
\centering
\includegraphics[scale = 0.7, trim = 0 0 0 0, clip]{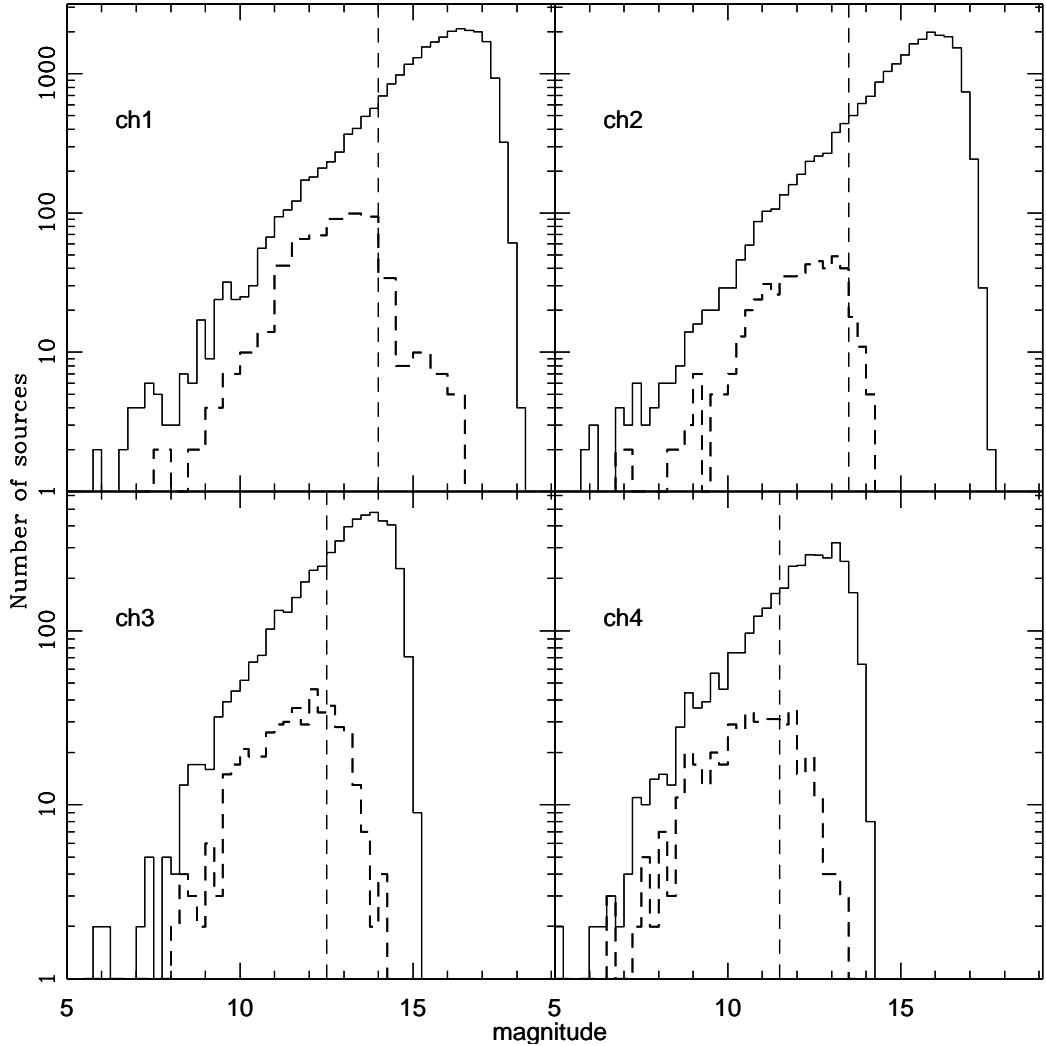}

\caption{Apparent magnitude histograms of detection in all the four IRAC bands having errors $<$ 0.2 mag are shown with
solid histograms  and dashed histograms are for the YSO detection. The  vertical dashed lines indicate 90\% completeness 
limit estimated from the artificial star method.  }
 
\label{comp}
\end{figure*}

To  evaluate the completeness of the census of IRAC detection
quantitatively, we plot histograms of the  IRAC  sources having error $<$ 0.2 mag for all
the channels and they are shown by using solid lines in Fig. \ref{comp}. 
In general, the data can be considered  complete till the linear distribution in the histograms 
 which are found to be  16.5, 16.0, 14.0 and 13.0 mag for 3.6, 4.5, 5.8 and 8.0 $\mu$m bands, 
respectively. However, the completeness of the YSO census is limited by many  factors. Bright
extended, variable  nebulosity in the IRAC bands can be found across the region, which significantly
limits the point source detection in these areas. The YSO identification from the 2MASS-
IRAC colours is limited by the modest sensitivity of the 2MASS survey, while that from
the IRAC-MIPS colours suffer from the significant saturation in the IRAC 8 $\mu$m and MIPS 24 $\mu$m images
caused by the central luminous sources as well as bright nebulosity.  Similarly the variable reddening (see Fig. \ref{extmap}) 
and stellar crowding characteristics across the extend of the region  (see Paper 1) 
could  also affect the local completeness limit. Completeness limits for the {\it Spitzer} photometry were also 
assessed by using a method of inserting artificial stars into the mosaics and then employing our
detection algorithms to identify them (e.g., \citealt{jose2008}). The sub-regions  A, B, C,  E and F of the Sh2-252 complex are found to be 
severely affected by the variable reddening and
nebulous background (Paper 1). In order to account for this variable background in the source detection, by following Paper 1,  
we estimated the completeness of two representative sub-regions i.e., regions A and C.  90\% 
completeness limits of these two regions have been obtained as 14.0, 13.5, 12.5 and 11.5 mag, respectively
for 3.6, 4.5, 5.8 and 8.0 $\mu$m bands and are shown by using vertical dashed lines in Fig. \ref{comp}.

The Sh2-252 complex lies at a distance of 2.4 kpc. The completeness limit of 14.0 mag at 3.6 $\mu$m 
corresponds to an approximate stellar mass of 1.0 M$_\odot$ for a YSO having  an  age of $\sim$ 1 Myr 
\citep{siess2000}. A more conservative estimate for the  90\% completeness of the PMS 
membership of the region can be made from the histogram distribution of candidate YSOs shown by using the
dashed lines in Fig. \ref{comp} which seems to be matching with the completeness estimation by artificial 
star method.  
%However, this estimate does not account for variable extinction, which will move completeness to even more higher masses. 
By taking variable extinction and  stellar crowding characteristics into account,  in conclusion, we consider 
that our  YSO census of the Sh2-252 complex is complete down to 1-2 M$_\odot$.

%%%%%%%%%%%%%%%%%%%%%%%%%%%%%%%%%%%%%%%%%%%%%%%%%%%%%%%%%%%%%%%%%%%%%%%%%%
%%%%%%%%%%%%%%%%%%%%%%%%%%%%%%%%%%%%%%%%%%%%%%%%%%%%%%%%%%%%%%%%%%%%%%%%  

\subsection{Characteristics of YSOs}

\subsubsection{From SED analysis}
\label{sed}

We constructed the  spectral energy distributions (SEDs) using the grid of models and fitting
tools of \citet{robitaille2006,robitaille2007} for characterizing and 
understanding the nature of  YSOs in the Sh2-252 complex. The models were
computed by \citet{robitaille2006} by using a Monte-Carlo based 20,000 2-D radiation
transfer models  from \citet{whitney2003a,whitney2003b,whitney2004} and by adopting several combinations of 
central star, disk, in-falling envelope, and bipolar cavity for a reasonably large
parameter space. Each YSO model has SEDs for 10 viewing angles (inclinations), so the
total YSO grid consists of 200,000 SEDs. The SED  fitting tool provides the evolutionary 
stage and physical parameters such as disk mass, disk accretion rate and stellar temperature 
of YSOs and hence is an ideal tool to study the evolutionary status of YSOs in star forming regions. 
%To constrain the YSO characteristics, we fit SEDs to those 183 sources (51 Class I, 132 Class II sources)
%identified in Sections \ref{yso_irac} and \ref{yso_2mass}  and having counterparts in 24 $\mu$m.
 Of the total YSOs identified in Sections \ref{yso_irac} and \ref{yso_2mass}, 183 sources (51 Class I, 132 Class II)
have counterparts in 24 $\mu$m and we fit SEDs to these 183 YSOs. The counterparts of these YSOs are searched in the optical
and NIR bands irrespective of their error criteria. Of these sources, 54 are found to  have counterparts
at least in two of the optical bands ($UBVRI$) along with the $JHK$, IRAC and 24 $\mu$m data. 96 sources
have $JHK$, IRAC and 24 $\mu$m data and rest of the sources  have IRAC and 24 $\mu$m data. 
Thus we have a minimum of 5  and maximum of  13 data  points in the wavelength range from 0.36 $\mu$m to 24 $\mu$m
to construct the SEDs. The SED fitting tool fits each of the models to the data allowing the distance and
external foreground extinction as free parameters.  We gave a distance range of 2.1 - 2.7 kpc 
(see Table 3 of Paper 1) as the input. An extinction range has been supplied as the input for each source from the $A_K$  
value estimated from the $(H-K)$ colour as discussed in Section \ref{kextmap} and by allowing  an 
uncertainty of $\pm$ 3 mag in the calculated extinction value. 
We further set photometric uncertainties of 10\% for the optical and NIR data, 15\%
for the IRAC and 24 $\mu$m data. These values are adopted instead of the formal errors in the catalog in order
to fit without any possible bias caused by underestimating the flux uncertainties. 
We obtained  the physical parameters of YSOs using the relative probability distribution for the stages 
of all the `well-fit' models. The well-fit models of each source are defined by

\begin{center}
$ \chi^2 - \chi^2_{min} \le 2N_{data}$,
\end{center}

where $\chi^2_{min}$ is the goodness-of-fit parameter for the best-fit model and $N_{data}$ is the number of
input observational data points. 

\begin{figure*}
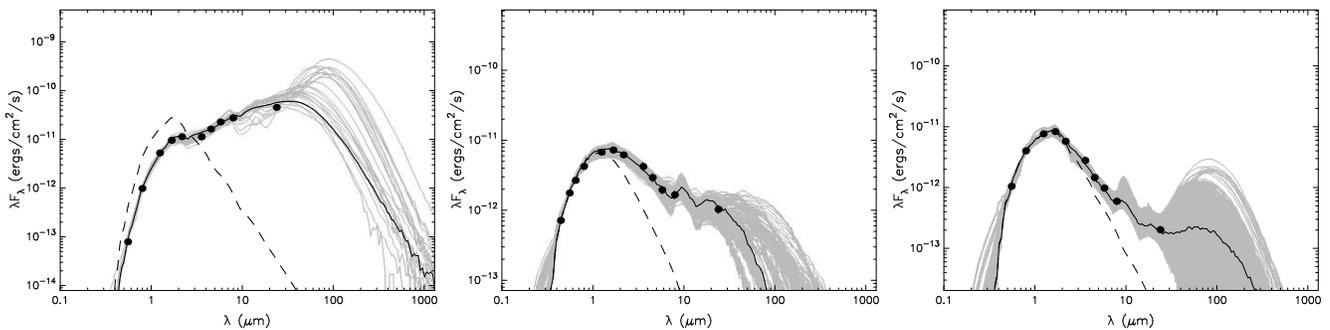
 \centering 
\includegraphics[scale = 0.49, trim = 0 0 0 00, clip]{fig6_a.eps}
\includegraphics[scale = 0.49, trim = 0 0 0 00, clip]{fig6_b.eps}
\includegraphics[scale = 0.49, trim = 0 0 0 00, clip]{fig6_c.eps}

\caption{Sample SEDs for a  Class I (left), Class II (middle) and transition disk (right) sources using the SED fitting tools of \citet{robitaille2007}. 
The solid black line shows  the best-fit and the grey lines show the subsequent well-fits. The dashed 
line shows the stellar photosphere corresponding to the central source of the best fitting model. The 
filled circles denote the input flux values. 
}

\label{seds}
\end{figure*}

In Fig. \ref{seds}, we  show the sample SEDs of  Class I, Class II and transition disk sources, where the solid
black line represents the best-fit and the grey lines are the subsequent well-fits. 
From the well-fit models for each source derived from the SED fitting tool, we calculated the $\chi^2$ 
weighted model parameters such as the stellar mass ($M_*$), temperature ($T_*$), stellar age ($t_*$), 
mass of the disk ($M_{disk}$), disk accretion rate ($\dot{M}_{disk}$),  envelope accretion rate ($\dot{M}_{env}$)
etc. The error in each parameter is calculated from the  
standard deviation of all well-fit parameters. The  parameters and the corresponding errors
of all the YSOs are listed in Table \ref{sedpars}. The IDs 1 - 51 and 52 -183 
represent Class I and Class II  sources, respectively. For some of the sources, 
the errors associated with few parameters (Table \ref{sedpars}) are quite large because 
we are dealing with a large number of parameter space, with limited number of observational data points.  
Additional observational data points in longer wavelengths would help  constrain these parameters more precisely. 

%%%%%%%%%%%%%%%%%%%%%%%%%%%%%%%%%%%%%%%%%%%%%%%%%%%%%%%%%%%%%%%%%%%%%%%%%%%%%%%%%%%%%%%%%%%%%%%%%
%%%%%%%%%%%%%%%%%%%%%%%%%%%%%%%%%%%%%%%%%%%%%%%%%%%%%%%%%%%%%%%%%%%%%%%%%%%%%%%%%%%%%%%%%%%%%%%%%

\begin{figure}
\centering
\includegraphics[scale = 0.6, trim = 35 30 30 120, clip]{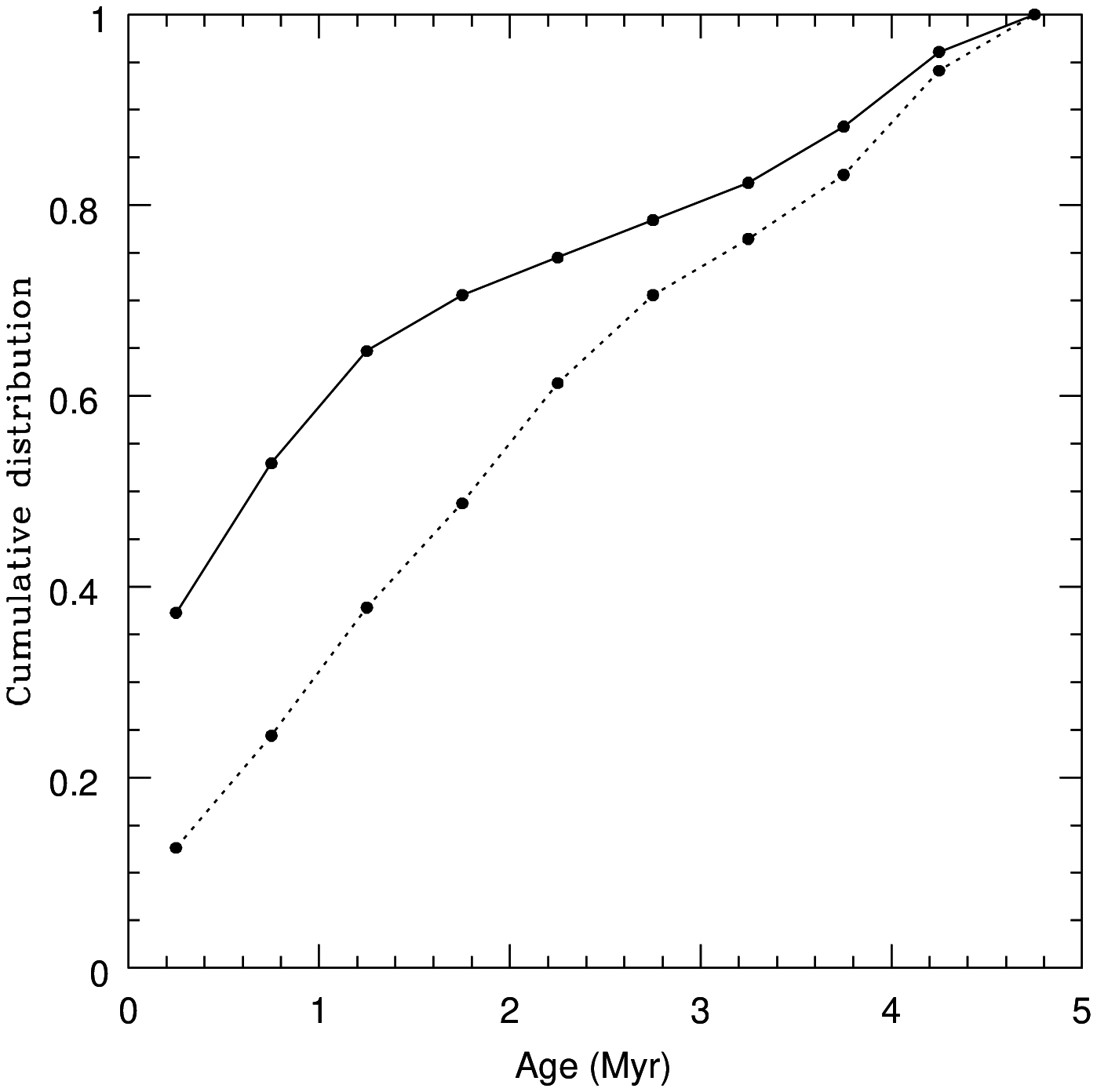}

\caption{Cumulative distribution of Class I (solid) and Class II (dotted)  YSOs as a function of age.} 
 
\label{cum}
\end{figure}

Table \ref{sedpars} shows that $\sim$ 94\% (172/183) of the sources have an age 
range between 0.1 to 5 Myr. This age distribution is in agreement with 
the age range discussed in Paper 1 using the optical colour magnitude diagram (CMD) of 
the candidate YSOs. Similarly, masses of the YSOs are distributed in a range between 
0.4 to 5.3 $M_\odot$ with a majority of them  having masses between 0.5 to 3.0  $M_\odot$. 
This is again in agreement with that of  Paper 1.  
%This shows that our  MIR photometry has a detection limit down to 0.4 $M_\odot$. 
The SED model parameters 
of YSOs, such as the  disk accretion rate and  envelope accretion rate are considered to be functions
of their  evolutionary status. Stage 0/I, II and III YSOs have significant infalling 
envelopes, optically thick disks and optically thin disks, respectively \citep{robitaille2006}.
In Table \ref{sedpars}, a majority of the Class II YSOs have the disk accretion rate of the 
order of $\sim$ $10^{-7}$  - 10$^{-8 }$  $M_{\odot}$  yr$^{-1}$. Similarly, the envelope
infall rates of most of the Class I YSOs are $>$ 10$^{-6 }$  $M_{\odot}$  yr$^{-1}$, which are 
relatively higher than that of the Class II YSOs, confirming their evolutionary stages. 
In Fig. \ref{cum} we have shown the cumulative distribution  of Class I and Class II YSOs as a function of their ages
which  manifests that Class I sources  are relatively younger than Class II sources. 
We have performed the Kolmogorov-Smirnov (KS) test for the age distribution of the Class I and Class II sources.
The test result shows that the chance of the two populations having been drawn from a same distribution is $<1$\%.
It suggests that the age distribution of two populations are different.

%%%%%%%%%%%%%%%%%%%%%%%%%%%%%%%%%%%%%%%%%%%%%%%%%%%%%%%%%%%%%%%%%%%%%%%%  

%%%%%%%%%%%%%%%%%%%%%%%%%%%%%%%%%%%%%%%%%%%%%%%%%%%%%%%%%%%%%%%%%%%%%%%%  

\subsubsection{From optical colour-magnitude diagram}
\label{cmd}

%%%%%%%%%%%%%%%%%%%%%%%%%%%%%%%%%%%%%%%%%%%%%%%%%%%%%%%%%%%%%%%%%%%%%%%%  

\begin{figure}
\centering
\includegraphics[scale = 0.6, trim = 35 30 30 120, clip]{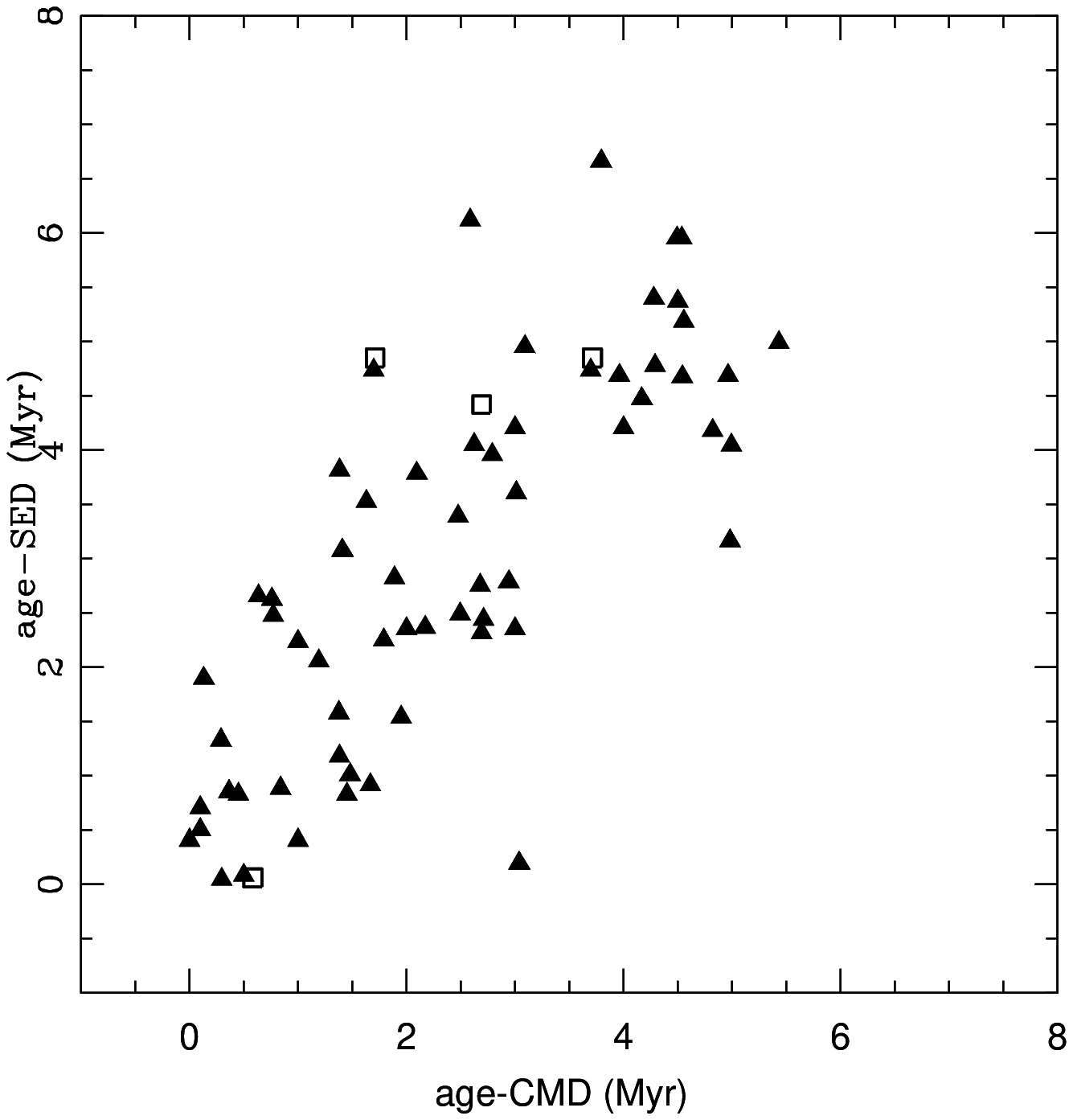}

\caption{Comparison of age distribution from CMD analysis with that from SED fitting. Squares and tringles represent the Class I and Class II YSOs, respectively} 
 
\label{compare}
\end{figure}
%%%%%%%%%%%%%%%%%%%%%%%%%%%%%%%%%%%%%%%%%%%%%%%%%%%%%%%%%%%%%%%%%%%%%%%%  

 The distribution of YSOs on the $V/(V-I)$ CMD is an ideal tool to estimate their approximate ages and masses.
The $V/(V-I)$ CMD for the candidate Class I, Class II and H$\alpha$ emission line sources are shown in the
Fig. 8 of Paper 1, which suggests that majority  of the YSOs in Sh2-252 are distributed between 
the PMS isochrones of age 0.1 and  5 Myr  and evolutionary tracks of  masses ranging from 0.5 to 3.0  $M_\odot$. This is in agreement 
with the SED based analysis (Section \ref{sed}).
% However, the photometric uncertainty, differential reddening,  binarity, different evolutionary status of YSOs
%etc. can cause spread in the CMD. 
The age of each YSO has been estimated from the CMD as discussed in \citet{pandey2008} and  
\citet{chauhan2009}. Briefly, their ages have been derived  by comparing their locations on the CMD with PMS isochrones of 
various ages \citep{siess2000}, after  correcting  for the distance and reddening of the region. 
Apart from the Class I, Class II and  H$\alpha$ sources, all other sources having age $<$ 5 Myr in the $V/(V-I)$ CMD and 
with $(H-K) >$  0.4 mag were suggested  as the candidate PMS members of the complex  in Paper 1.  We also estimated the
 ages of these additional candidate PMS sources as discussed above. There are 80 YSOs which have age estimation both from 
the CMD  and SED (Section \ref{sed}) analyses.   In Fig. \ref{compare} we have shown the comparison between the age distribution from
CMD analysis with that from SED fitting.  Squares and triangles represent the Class I and Class II YSOs, respectively. 
Though the age estimation from both the methods are quite scattered, in general, the distribution can be considered correlated with each other. 
 We have also performed the KS test between the age distribution of YSOs  from SED
and CMD analyses and  found that the ages estimated by these two methods are correlated by a factor of $\sim$ 81\%. 
Hence, we confirm that our age estimation based on both methods is in agreement.

\subsection{The ages of sub-regions in Sh2-252}
\begin{figure*} \centering
\includegraphics[scale = 0.6, trim = 0 0 0 0, clip]{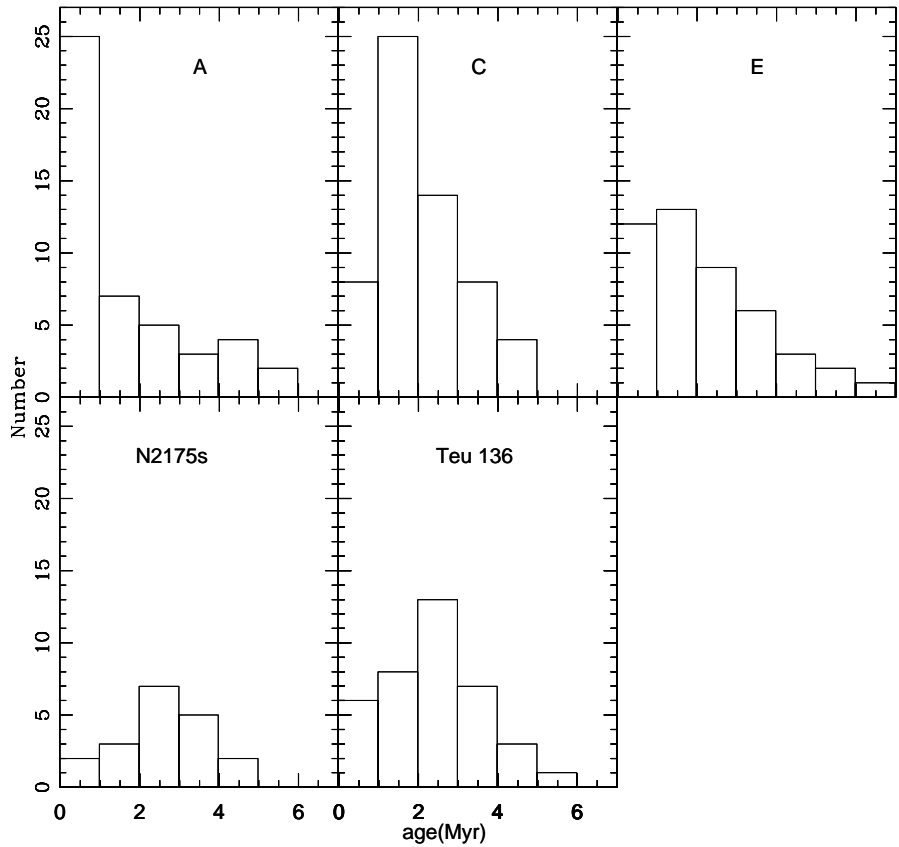} 
\caption{Histograms of  age distribution estimated from the SED and CMD analyses for candidate  YSOs within the five sub-regions of Sh2-252 complex with
median age of 0.5, 1.5, 1.5, 2.5 and 2.5 Myr, respectively for the regions A, C, E, NGC 2175s and Teu 136. } 

\label{hist}
\end{figure*}

 We attempted to constrain the  evolutionary status of the sub-regions of Sh2-252  using the SED fitting  and CMD analyses. 
Here we would like to caution that 
although the age estimation using the SED fittings  may not be robust,  the results agree reasonably well with the 
ages estimated by us using $V/(V-I)$ CMD for a subset of YSOs (Section \ref{cmd}).   Hence we accept  the ages 
determined by models with reasonable confidence.  
% The surface stellar density analysis in $K$-band (Paper 1) shows that the Sh2-252 complex has 
% at least five levels of prominent  stellar density enhancements associated with the sub-regions A, C, E, NGC 2175s and Teu136. 
We estimated the ages of the sub-regions A, C, E, NGC 2175s and Teu 136 of Sh2-252 by using the ages of the candidate YSOs  
lying within their boundaries.   The boundaries of these individual regions have been taken from the  stellar surface density map. i.e., 
3$^\prime$.5, 3$^\prime$.5, 3$^\prime$.0, 3$^\prime$.0 and  3$^\prime$.0, respectively for regions A, C, E, NGC 2175s and Teu 136 (Paper 1).  
The histograms of age  distribution estimated from the SED and CMD analyses of each sub-region are shown 
in Fig. \ref{hist}.  The median ages of YSOs within A, C, E, NGC 2175s and Teu 136 have been obtained as $\sim$ 0.5, 1.5, 1.5, 
2.5 and  2.5 Myr, respectively. As evident in the $K$-band surface stellar density map (Paper 1), 
no clustering is apparently seen towards the region B. From the SED  analysis of 3 candidate YSOs lying  within 2$^\prime$.5 
radius of region B, a  median age of 1-2 Myr can be  assigned  for this region. However, the age  of Sh2-252 B should  be 
considered as an approximate value since the statistics is very poor. The 
above age estimates  based on the SED/CMD analyses show that  the region A is the youngest among all the sub-regions, whereas regions   B, C and  E  
are of similar age and  NGC 2175s and Teu 136 are found to be slightly evolved than other sub-regions.  

%%%%%%%%%%%%%%%%%%%%%%%%%%%%%%%%%%%%%%%%%%%%%%%%%%%%%%%%%%%%%%%%%%%%%%%%%%
%\begin{landscape}
%\end{landscape}
%%%%%%%%%%%%%%%%%%%%%%%%%%%%%%%%%%%%%%%%%%%%%%%%%%%%%%%%%%%%%%%%%%%%%%%%  

\subsection{Spatial distribution of YSOs in Sh2-252}
\label{spatialdist}

\begin{landscape}
\begin{figure}
\centering
\includegraphics[scale = 1.5, trim = 40 25 20 270, clip]{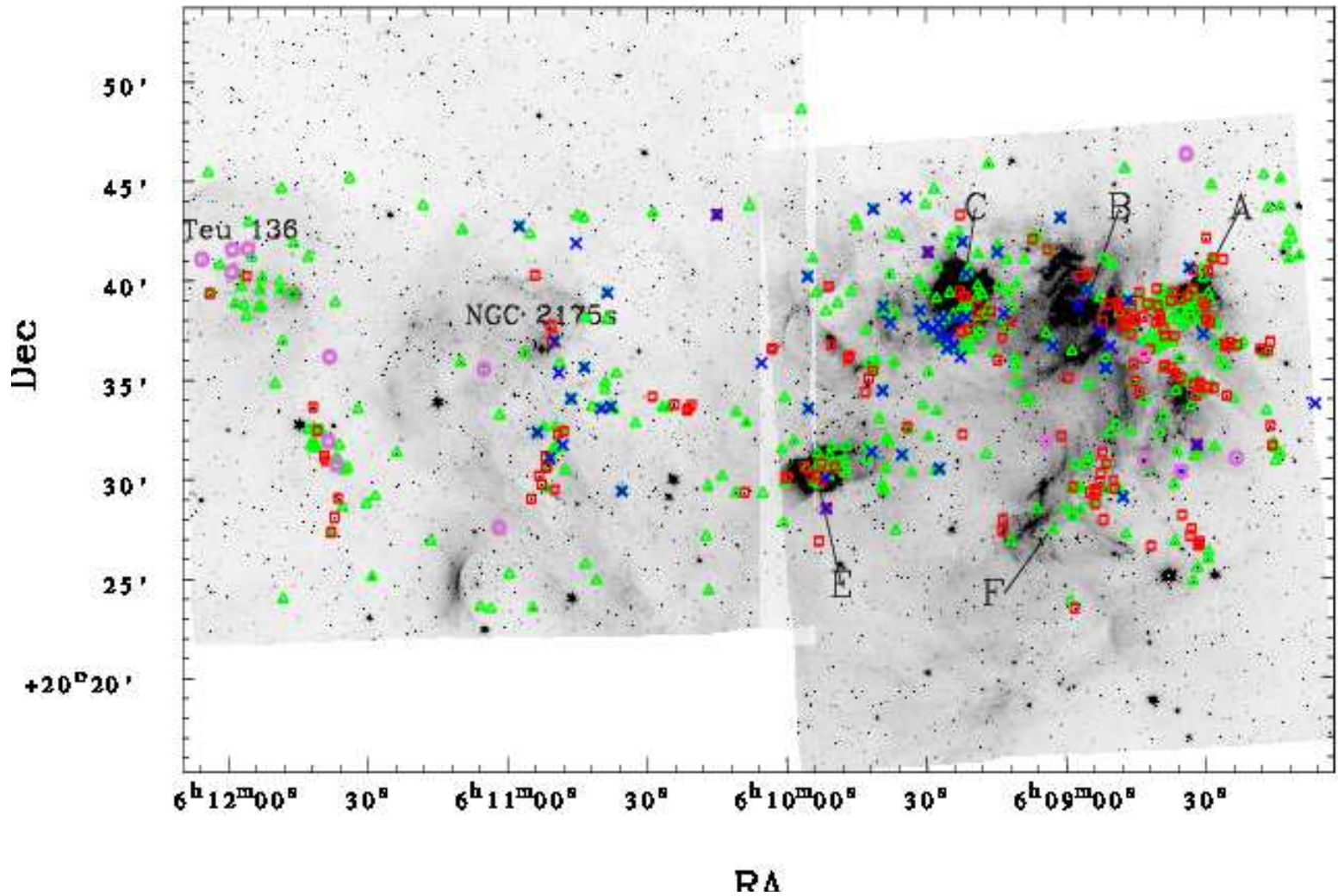}
\caption{Spatial distribution of  candidate YSOs in Sh2-252 over plotted on the 
4.5 $\mu$m mosaic image. The candidate Class I, Class II and transition disk  sources are shown by using red squares, green triangles and
magenta circles, respectively and the blue crosses  are the $H\alpha$ emission line sources identified 
from the slitless spectroscopy survey.  The main sub-regions of Sh2-252  are  marked in the figure  and the 
star symbol represents the location of HD 42088. RA and Dec are in J2000 epoch (see the electronic version for the colour image).  }
\label{spatial}
\end{figure}
\end{landscape}

%By examining the spatial distribution of the young stars in a given star forming regon, it is 
%possible to trace the underlying structure of the molecular cloud from which they form and then to
%understand its star formation history.
Spatial distribution of YSOs in a star-forming complex helps to trace  its star formation history.
In Fig. \ref{spatial}, we  show the spatial distribution of all the identified candidate YSOs in
Sh2-252 (i.e., red squares: Class I; green triangles: Class II; magenta circles: transition disk sources)  overlaid on the 
IRAC 4.5 $\mu$m mosaic image.  The 61 $H{\alpha}$ emission line sources 
identified from slitless spectroscopy survey (Paper 1) are shown by using blue crosses. The important sub-regions
of Sh2-252 are also marked in the figure. This figure indicates some interesting spatial sequence of YSOs
within Sh2-252.  A comparison of this figure with Fig. \ref{color-irac} clearly indicates that a 
majority of the candidate YSOs identified  in Sh2-252 are preferentially concentrated around the C\hii regions A, 
C and E, respectively. Also,  the bright rim feature discussed in Section \ref{intro} (i.e., region F), which 
makes up the western border of the complex contains several number of YSOs.  
There are not many YSOs detected towards the region B. 
%The locations of these YSOs %can be easily correlated with the WCF and ECF identified by Lada \& Wooden (1979) (see Section \ref{intro}).  
The WCF is seen to have more number of YSOs when compared to the ECF. Overall, the distribution of YSOs 
in the west of Sh2-252 shows a nice correlation with the semi-circular shaped  distribution of  molecular gas 
and the PAH emission/\hii region cavity boundaries (see Section \ref{morphology}). The strong positional coincidence 
between the YSOs and the molecular cloud towards the WCF in Sh2-252 shows the enhanced star formation activity towards 
this region as observed in other star forming regions (e.g. \citealt{evans2009}, \citealt{fang2009}). There are a few number of 
sources scattered towards the east of Sh2-252. There are two groups of YSOs aligned in a filamentary manner
associated with the cloud remnants seen towards  the east of C\hii region E  (see Figs. \ref{area} and \ref{extmap}). However,
the cluster NGC 2175s does not seem to  have significant number of YSOs.   There seems to be a small clustering of 
Class II and transition disk YSOs  towards Teu 136.  It is also interesting to note that there is an enhanced  concentration of 
Class I YSOs towards the east of the sub-region A  when compared to other sub-regions of Sh2-252.  
This  is an indication that the region A  might be at a younger evolutionary  stage and we will be iscussing 
more on this in the ensuing sections. 

 The number of YSOs in each class is useful for estimating the relative age of a star forming region (\citealt{hatchell2007}; \citealt{gutermuth2009}).
Since Class II sources are assumed to be more evolved than  Class I and hence the Class I/II ratio is a function of the 
evolutionary status of a system \citep{beerer2010}.  We used this ratio  as a proxy for the relative ages of the 
sub-regions in Sh2-252. The spatial 
distribution of the candidate YSOs given in Fig. \ref{spatial} shows that most of them are distributed  within 
or in the close proximity of the sub-regions  A, C, E  and Teu 136.  Hence we estimated the ratio of Class I to 
Class II YSOs for these sub-regions  within their  radius (Paper 1). The  Class I/II ratios 
have been obtained as 0.54, 0.24, 0.36 and 0.13, respectively  for 
regions A, C, E and Teu 136. This fraction is  of almost  similar order 
%as that in well studied star forming regions  such as ChaII (0.16),
%Lupus (0.29), Serpens (0.42) and Ophiuchus (0.47) \citep{evans2009} and 
with the median ratio of Class I to Class II sources (0.27) obtained in nearby clusters 
\citep{gutermuth2009}.  The various sub-regions within the S254-258 complex, a nearby star forming region to Sh2-252 in the Gem OB1 association, 
also have the Class I/II ratios vary between
0.32 to 0.78 \citep{chavarria2008}.  Thus the  Class I/II ratios of the sub-regions of Sh2-252 can be considered comparable with the other star forming regions.  
In this scheme, the high concentration of Class I YSOs towards the  region A  suggests that the 
youngest population is  located towards the western boundary  of the complex.

%%%%%%%%%%%%%%%%%%%%%%%%%%%%%%%%%%%%%%%%%%%%%%%%%%%%%%%%%%%%%%%%%%%%%%%%%%
%%%%%%%%%%%%%%%%%%%%%%%%%%%%%%%%%%%%%%%%%%%%%%%%%%%%%%%%%%%%%%%%%%%%%%%%

%This fraction is  of almost similar order as that in well studied star forming regions  such as ChaII (16\%),
%Lupus (29\%), Serpens (42\%) and Ophiuchus (47\%) (Evans et al. 2009).

%\subsection{A candidate runaway star in the system}
%\input{runaway}

\section{Discussion}
\subsection{Probable evolutionary status of the sub-regions in Sh2-252}
\label{evolution}
Since most of the giant molecular clouds (GMCs) in the Galaxy are hierarchical in structure, the evolutionary  
status of the identified YSOs helps to get a deep insight about the overall star formation activity of the complex. 
\citet{tej2006} studied the sub-region  A using  NIR, sub-mm  and radio observations. Based on their $K$-band 
surface density map, which is  deeper and of  better resolution than ours, they found sub-structures
within this region. In \citet{tej2006}, the prominent peak of the  stellar density coincides spatially 
with the central ionizing source of A, which is 
an O9.5V type star (Paper 1). However, a secondary stellar density peak, which lies $\sim$ 50$^{\prime\prime}$ east to the main peak has 
also been found and its spatial location coincides with  the sub-mm peak detected at 850 $\mu$m \citep{tej2006}. The authors 
have attributed this as a possible  cold core at an early evolutionary stage than A. Similarly, the most intense CO peak of the WCF 
is also located east  to  the region A with  water and methanol maser emissions (\citealt{lada1979}; 
\citealt{kompe1989}; \citealt{szymczak2000}) within its  proximity, suggesting the recent  star formation
activity towards the east of A (\citealt{lada1979}; \citealt{lada1981}; \citealt{kompe1989}).  The water and methanol maser emissions 
associated with 850 $\mu$m and CO peak towards the  east of  region A suggests that this region  is   possibly at an  early evolutionary stage than A.

The high concentration of Class I YSOs around the region A (Fig. \ref{spatial}) and the large value of Class I/II ratio  
(see Section \ref{spatialdist}) show that  region A is at a  young  evolutionary stage compared to the other sub-regions of the complex. 
The average age estimation ($\sim$ 0.5 Myr) based on the SED and
CMD analyses also shows that the region A is indeed  very young. Based on the class I/II ratio  and age histogram analysis,
the regions B, C and E are found to be at a similar evolutionary stage and are slightly evolved (median age $\sim$ 1-2 Myr) than region A.    
The presence of a very few number of YSOs
towards  the small cluster NGC 2175s (see Fig. \ref{spatial}) suggests that either the  cluster was formed in an early epoch of star formation
in Sh2-252 or it is a case of rapid circumstellar disk evolution of YSOs  (e.g., \citealt{bally1998}) due to the presence of 
four early type massive sources  at its center (see Table 4 of Paper 1). 
%The strong stellar winds from the massive stars  would have 
%caused an early  disk evaporation  in NGC 2175s (e.g., \cite{bally1998}). 
However, the median age  estimation based on the CMD analysis gives an approximate age of $\sim$ 2-3 Myr for this cluster. The small 
cluster Teu 136 lying towards the east of Sh2-252 seems to have more number of 
Class II YSOs than Class I. The low value of the class ratio  as well as the SED/CMD based age analyses (median age $\sim$ 2-3 Myr)
show that this cluster is also at a slightly more evolved stage when compared to the sub-regions  A, B, C, and E of the Sh2-252 complex.

\subsection{Star formation activity towards WCF-The general picture}

Sh2-252 is a large \hii region in the Gem OB1 association, excited by the O6.5V star HD 42088 of age 2-4 Myr. 
It contains 12 B-type stars, some of them are associated with  C\hii regions. Since C\hii regions are at a early stage
of massive star formation, their impact is limited to their local environment. Thus,  O6.5V star is the most
dominant source of energy responsible for ionization in the region.  \citet{lada1979} showed  that the cloud
complex is separated  mainly into two fragments ECF and WCF, which themselves are embedded in a 
more diffuse and extended ($\sim$ 90 pc) molecular cloud complex.   Most of the mass of WCF ($\sim$ 2.5 $\times$ 10$^4$ \msun) 
is concentrated in a narrow ridge, which borders the western IF of Sh2-252.  This  dense, semi-circular shaped shell 
observed at the eastern boundary of WCF, appears to have been swept up by the pressure from the IF generated by the 
exciting source of Sh2-252 \citep{lada1979}. 
%These components giving morphological appearances that they might have been swept 
%up by the expanding IF and/or the stellar winds of the ionization source of the \hii region. 
%Shells around \hii regions are potential sites of  subsequent star formation. 
%owing to the  energy inputs from the nearby massive stars. 
%We observed many YSOs in this
%direction preferentially distributed at the western outskirts of the \hii region (see Fig. \ref{spatial}). 
The  large size and massive content of WCF suggest that
the entire cloud  is unlikely to be swept up. We roughly compared  the required kinetic energy
and the energy produced by the massive O6.5V star to sweep up the entire shell. 
The combined energy (radiation plus wind) liberated by the O6.5V stars is found to be less than the 
required energy to drive the  entire cloud (WCF+ECF $\sim$ 2.7$\times$10$^4$ \msun; \citealt{lada1979}).
 Thus it is  suggested that at least part of the shell in WCF is probably primordial in nature and the
entire shell is not swept up by the expansion of the \hii region. 
However,  at the same time  it is obvious from  Fig. \ref{color-irac} that  the \hii region is in interaction with WCF, with bright 
semi-circular PDR and compressed neutral material behind it and protruding structures such as `finger tips'
and a bright rim pointing towards the ionizing source. Moreover, we observed recent star formation activity
towards this region. The morphology resembles that star formation  has occurred in thin shells around the \hii region  
and  star formation seems to be more active in this direction compared to the whole cloud  complex.

%%%%%%%%%%%%%%%%%%%%%%%%%%%%%%%%%%%%%%%%%%%%%%%%%%%%%%%%%%%%%%%%%%%%%%%%%%
\begin{figure*} 
\centering
\includegraphics[scale = 0.57, trim = 0 0 0 0,   clip]{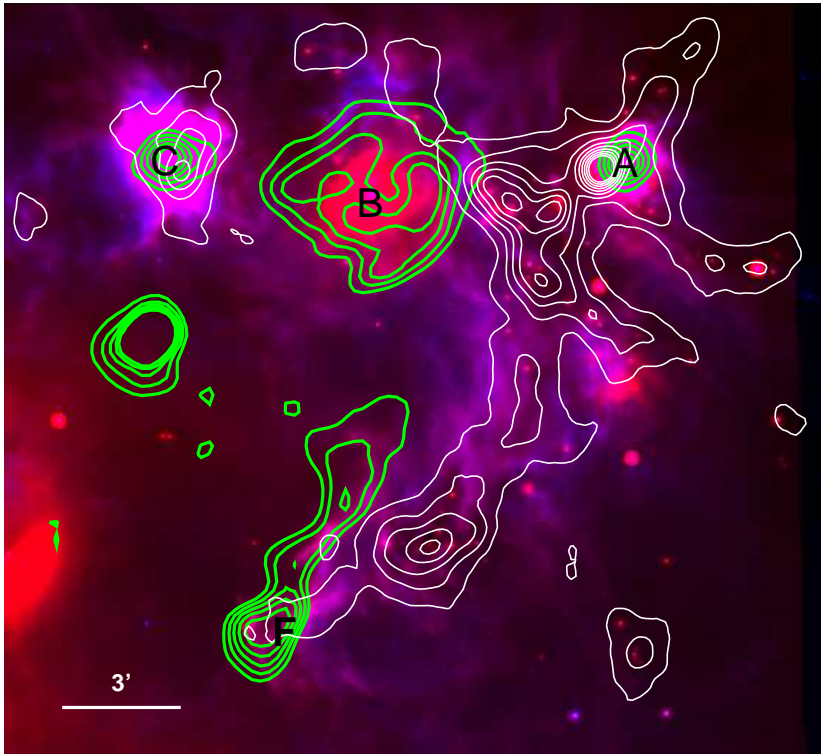}
\caption{ Colour composite image of the western half of Sh2-252 made from the 8.0 and 24 $\mu$m  bands along with the
1.1 mm dust continuum emission map (white;   contour levels at 0.06, 0.21, 0.37, 0.52, 0.68, 0.83, 0.99, 1.14 and 1.30 Jy/beam) 
and the low resolution radio continuum map at 1280 MHz (green;  contour levels at  0.25, 0.44, 0.63, 0.82, 1.02 mJy/beam). 
The sub-regions are marked in the figure. North is up and east is to the left.}
\label{color-24}
\end{figure*}
%%%%%%%%%%%%%%%%%%%%%%%%%%%%%%%%%%%%%%%%%%%%%%%%%%%%%%%%%%%%%%%%%%%%%%%%  
 Fig. \ref{color-24} shows the  colour composite image of the western half of Sh2-252 made from the 8.0 and 24 $\mu$m bands. 
The contours of 1.1 mm dust continuum  emission (taken from Bolocam Galactic Plane Survey; \citealt{aquirre2011}) and low resolution 
radio continuum emission at 1280 MHz (Samal et al. in preparation) are over plotted  in Fig. \ref{color-24}.
 On a large scale we see a ring of molecular material at 1.1 mm aligned parallel to the outer extent of the 
          IF of Sh2-252 seen in  optical (see Fig. \ref{area}), and  located just behind the bright PAH emission seen in 
          8 $\mu$m and also corresponds
          to the dark region at 24 $\mu$m. The molecular material in the ring seems to be fragmented at various locations. This shell of
          molecular material located behind the  IF  provides  evidence for the  most compressed zone of the WCF fragment. We also observed
          many Class II and Class I YSOs in this direction (Fig. \ref{spatial}), with a majority of them projected behind the PDR, in the
          vicinity of 1.1 mm emission. Apart from the YSOs,  three C\hii regions (A, B and C) are observed in the close
          vicinity of the 8 $\mu$m ring, with region A being located slightly behind the PAH ring and separated from the
          ring by a dark lane, consisting of highly dense molecular material.  The region A is part of the 
        massive (M $\sim$ 2.5$\times$10$^4$ $\msun$), large ($\sim$ 16 pc) WCF cloud  (see Fig. \ref{area}).
          Hence the influence of O6.5V star to region A, if at all exists,   must be weak. 

         %   The ages based on associated YSOs within region B \& C is 1-2 Myr, similar to
         % those found for the YSOs associated with region E. Though, the region E is located in the very close
         % proximity of the ionizing O6.5 V star,  its IR morphology does not seem to be strongly affected  by the
         % strong stellar radiation and wind from the star. This indicates that  it is possibly projected along the
         % line of sight and  in reality, it  might be located away from the ionizing  star as those C\hii regions B and C. 
        {%\bf Ignoring the projected location of region E, 
       In general, the   configuration (at least in the western direction) 
       suggests that the \hii region has collected some of the matter during its expansion at its  periphery, like the  
       collect \& collapse process of star formation such as in Sh2-212 (\citealt{deharveng2008}; \citealt{jose2011}).}  
       Collect \& collapse process works well in homogeneous and uniform medium, which is definitely not the case here,  
    with openings at various directions  and without a clear circular shell structure. However it 
      has been found that collect \& collapse process can still be applicable in turbulent medium \citep{deharveng2008}.  From 
      the extinction map (Fig. \ref{extmap}) we found that the visual extinction  in the direction of the PAH ring 
        is, $A_V$ $>$ 5.5  mag, which  corresponds to a column density of hydrogen, $N(\mathrm H_2$) in the region 
         $>$ 1.03$\times$10$^{22}$  atoms cm$^{-2}$,   using the relation  $A_V$  = 5.34$\times$10$^{-22}~N(\mathrm H_2)$,
         derived from \citet{bohlin1978} for $A_V$ = 3.1$\times$E(B-V). The observed column  density towards the 
         molecular ring is thus comparable to the  column density required (i.e., $\sim$ 6$\times$10$^{21}$  atoms cm$^{-2}$) 
         in the collected layer for the fragmentation to happen, as predicted by \citet{whitworth1994}. 
         Assuming the observed star formation
         near the vicinity of the IF might have happened due to collect \& collapse process, we compared the observed properties  
        with the predication by the  numerical simulation of \citet{whitworth1994}. We adopted 1.7$\times$10$^{49}$ 
        as ionizing photons s$^{-1}$ (for O6.5V star; \citealt{vacca1996}), the isothermal sound speed as 0.2-0.3 \kms and 
        the density of the neutral material into which the  \hii region evolved as 400 cm$^{-3}$ \citep{lada1979}. With these
        values, the predicted  fragmentation time and radius at which fragmentation would take place are found to be 2.3-2.9 Myr and 
        10.3-11.6 pc, respectively. Though there are large uncertainties involved in these calculations, 
        considering the age of the ionizing star is of  $\sim$ 2-4 Myr (Paper 1) and  ages of the Class I/II YSOs  projected on the shell 
       $\sim$ 0.5 Myr, as well as the fact that  the farthest separation of the shell from
        the ionizing star is $\sim$ 10  pc, the star formation activity could in fact might have taken place  in the   collected shell. 
       
        The possible formation of regions B, C and E is in doubt by the same scenario, owing to their large age ($\sim$ 1-2 Myr; Section \ref{evolution}). 
        These objects might have  formed before the ionization-shock front passed over them, although with
        the present data it is hard to prove  whether the passage of ionization-shock front makes the age of these sources look
        older. Ignoring this fact, we suspect  that they are possibly not triggered due to the  expanding \hii region, and 
       more likely to  be primordial in nature, resulted from the formation and evolution of the molecular cloud itself, which is in agreement with
       \citet{lada1979}.

%%%%%%%%%%%%%%%%%%%%%%%%%%%%%%%%%%%%%%%%%%%%%%%%%%%%%%%%%%%%%%%%%%%%%%%%%%
\begin{figure}
\centering
\includegraphics[scale = 0.45, trim = 0 0 	0 0,   clip]{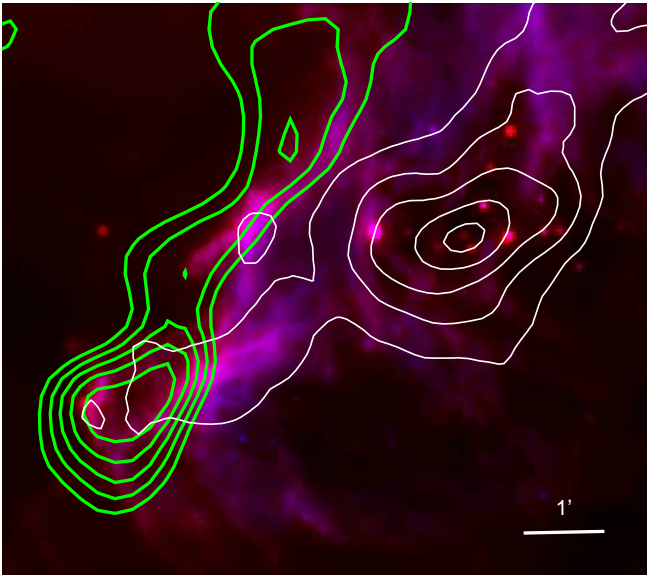}
\caption{Close-up colour composite image of region F made from the 8.0 and 24 $\mu$m  bands along with the
1.1 mm dust continuum emission map (white) and the low resolution  continuum map at 1280 MHz (green).
North is up and east is to the left.}
\label{color_close}
\end{figure}
%%%%%%%%%%%%%%%%%%%%%%%%%%%%%%%%%%%%%%%%%%%%%%%%%%%%%%%%%%%%%%%%%%%%%%%%  

\begin{figure*}
\centering
\includegraphics[scale = 0.8, trim = 2 2 0 2, clip]{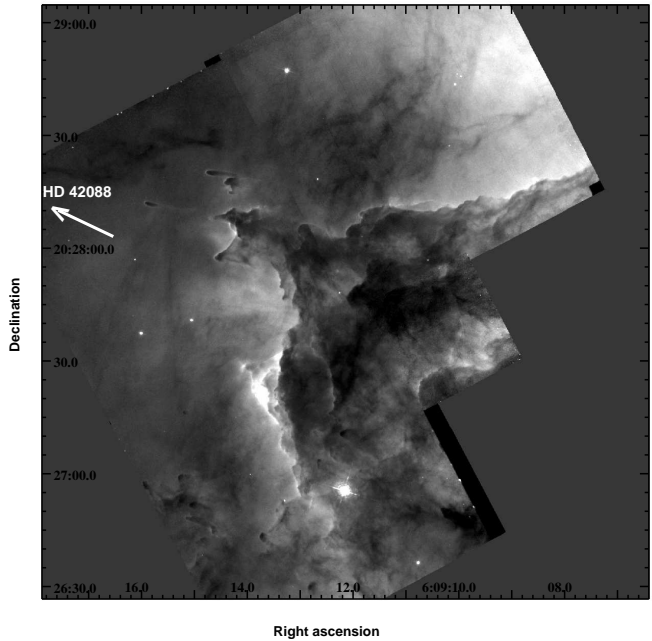}

\caption{HST-WFPC2 archive image at 0.65 $\mu$m of  the sharp edge of the molecular ridge in region F. The direction of the ionizing
source HD 42088 is  marked in the figure.  }

\label{hst}
\end{figure*}

 The radio source F is an  extended one, with no internal heating source. The radio emission is 
located ahead of the 8 $\mu $m emission (see Fig. \ref{color_close} for a close up view of region F) and corresponds to the glowing gas 
in the optical image (see Fig. 2 of Paper 1), thus, more likely due to  the 
photo-evaporating  gas from the surface of the cloud exposed to radiation.
This phenomenon usually occurs when a massive star drives an ionization-shock front into the ISM,
the external layers  of the cloud  exposed  to radiation starts photo-evaporating, and can be 
seen as glowing layer of gas. In Fig. \ref{hst}, the HST-WFPC2 archive image at 0.65 $\mu$m  
(Program No: 9091; PI: J. Hester) of the extreme edge of the  molecular ridge at region F
(i.e., the brightest portion of the bright-rim feature in Fig. 1 of Paper 1) is shown.
  Apart from the photo-evaporating gas,  this better resolution  image  displays number of `finger-tips and pillars' like fragments which are seen
silhouetted against the bright background pointing  towards the direction of HD 42088. 
%Such globules and finger-tips like structures can  also be formed via hydro dynamical instability 
%during the expansion of the \hii region without pre-existing molecular clumps  as those simulated by
%\cite{garcia1996}. The observational evidences of such regions has been
%found in W5 cloud complex and are mainly associated with low mass stars (see e.g., \citealt{chauhan2011}). 
%\cite{chauhan2011} suggested that  stars can be formed on such hydro dynamical instability structures 
%under the compressing effects of the \hii gas. 
 Three YSOs are observed at the tip of region F (see Fig. \ref{spatial}) and they are low mass ($<$ 2\msun)  Class I/II  sources,
 thus possibly formed from low mass clouds. Assuming that the  impact of ionizing radiation to such 
cloud structures initiates new star formation  similar to those in BRCs 
and small globules, we compared the observed incident ionizing flux to the required range of 
ionizing fluxes that can trigger star formation in fiducial cloud as predicted  by  \citet{bisbas2011} using numerical simulations.
\citet{bisbas2011} numerically 
estimated the required ionizing flux range as 10$^9$ cm$^{-2}$ s$^{-1}$ -  3$\times$10$^{11}$ cm$^{-2}$ 
s$^{-1}$. They predicted that if the ionizing flux is very low ($<$10$^9$ cm$^{-2}$ s$^{-1}$) then the cloud 
will be slowly eroded and if the ionizing flux is high ($>$10$^{11}$ cm$^{-2}$ s$^{-1}$) then the  cloud 
will be rapidly dispersed, with no star formation. We calculated  the ionizing photon flux 
impinging onto  these structures  from the radio emission from the
surface layer of the cloud following \citet{lefloch1997} assuming the 
electron temperature of the photo-evaporating gas as 10$^4$ K, and it  is estimated to be  3$\times$10$^9$ cm$^{-2}$ s$^{-1}$. 
The observed estimation is well within the predicted range to initiate star formation. 

Besides the low-mass YSOs at the tip of the pillars, we also observed a  clear clustering or concentration
of YSOs in the dense region of the cloud located behind region F (see Figs. \ref{color-irac} and \ref{spatial}).  The age of
these YSOs is of the order of $\sim$ 1 Myr, thus indicates the youth of the region.
These  YSOs seem to have resulted from the gravitational fragmentation of
the collected material rather than the implosion of pre-existing clumps due to 
external pressure. The YSOs show a chain-like elongated
distribution parallel to the IF and appear to lie in  the
collected shell. Whereas in the case of radiation driven implosion (RDI), the YSOs are expected to 
lie perpendicular to the direction of progressive IF, with younger YSOs being  located 
away from the IF. The incident photon flux is also on  the lower-side of the flux 
needed to initiate  RDI, thus possibly is not enough for a massive cloud to  form a  cluster of stars. 
Hence, we suspect that  the star formation is  more likely due to 
gravitational instability in the collected material, such as seen in the 
massive condensation of RCW120, where a chain of  YSOs is observed parallel to the IF \citep{deharveng2009}.

In Fig. \ref{spatial} we find the  densest concentration of YSOs  in the close vicinity of the molecular clump
located between the C\hii regions A and B (see Fig. 15 of \citealt{tej2006}), with many YSOs of 
Class I nature ($\sim$ 10$^5$ yr). This cloud clump is associated with  water and methanol masers 
and does not show any radio emission (down to 0.4 mJy). Hence, \citet{tej2006} suggested that 
it could be  a site of high-mass star-forming protocluster in a very early evolutionary stage (Section \ref{evolution}). We also observed
a cluster of YSOs, mostly Class I (Fig. \ref{spatial}), thus reflecting it is indeed a 
site of cluster formation sandwiched between the  two  relatively evolved C\hii regions A and B.   
Using H$_2$ observations  at 2.12 $\mu$m, \citet{tej2006}  traced a cometary-arc shaped shocked molecular H$_2$ emission 
bounding the ionized gas (i.e., ionization bounded side of the C\hii region A) in the direction 
of the molecular clump, possibly reflecting the interaction of the \hii region with the adjacent molecular cloud. 
Similarly, we noticed sharp radio contours for region B, in the direction of the molecular clump (Fig. \ref{color-24}). 
The radio size of  region B is larger than regions A, C and E, and is ionized 
by a similar kind of star (B0-B0.5 type; Paper 1).  Thus, if they are evolved in a cloud clump of  similar density, the larger size
of region B possibly reflects that it is relatively evolved in nature. The region between A and B  indeed reflects a
site of recent star formation, where much younger sources are found in the matter in between the two
relatively evolved C\hii regions A and B. A similar kind of star formation has been observed at the interfaces of
two expanding super giant shells LMC4 and LMC5 \citep{cohen2003} and two classical \hii regions S255-S257 
\citep{ojha2011}, where it is believed that collision of dense swept up 
neutral materials associated with individual shells lead to star formation 
(\citealt{yamaguchi2001a,yamaguchi2001b}). Possibly, we are witnessing  one such region, where star formation has been enhanced
due to a similar kind of process.

%at 6cm, for region A, a sharp edge is observed at it east by felli et al. 1977
%Components A, B, C aooear to be individual and independent Hii regions of higher surface brightnes, probably excited internally by stars of spespectral type later than )6.%. they r located near the edge of the extended source, component A being the more distant one. 

%%%%%%%%%%%%%%%%%%%%%%%%%%%%%%%%%%%%%%%%%%%%%%%%%%%%%%%%%%%%%%%%%%%%%%%%%%
%%%%%%%%%%%%%%%%%%%%%%%%%%%%%%%%%%%%%%%%%%%%%%%%%%%%%%%%%%%%%%%%%%%%%%%%
The overall picture emerges from our analyses is that the star formation activity observed towards WCF is 
more likely multi-fold and the  sources are found to be at different evolutionary stages.
Though the large scale CO map \citep{lada1979} provides much information about the global structure of the
complex, but the region certainly deserves attention for high resolution molecular line observations to explore the
detailed properties of the sub-regions.

\section{Summary}
The optically bright, extended \hii region Sh2-252 is a part of the Gemini OB1 association. This region is  mainly composed of  two small clusters
NGC 2175s and Teu 136 and four C\hii regions namely A, B, C and E. In this paper, an extensive survey of the star forming complex 
Sh2-252 has been undertaken with an aim to explore its hidden young stellar population, their characteristics, spatial distribution,
morphology of the region and finally to understand the  star formation scenario of the complex for the first time. 
{\it Spitzer}-IRAC, MIPS photometry (3.6 - 24 $\mu$m) are combined with 2MASS-NIR and optical data sets to identify and classify the 
YSOs by their IR excess emission from their circumstellar material. Using the IR C-C criteria, we have identified 577 YSOs 
in the complex,  of which, 163 are consistent with Class I, 400 are consistent with Class II and 14 are consistent with transition 
disk YSOs, suggesting a moderately rich number of YSOs in this \hii region. From the  CMD and SED based analyses, majority of the YSOs 
are found to have an age distribution between 0.1 - 5 Myr and mass in the range of 0.3 - 3.0 M$_{\odot}$.

Spatial distribution of the candidate YSOs shows that they  are mostly clustered around the sub-regions of the complex such as in A, C, E, F 
and Teu 136. However, majority of the candidate YSOs are distributed in the western part of the complex when compared to the east, suggesting 
enhanced star formation activity towards its west. Using the SED and CMD  based age analyses, we derived probable evolutionary status of the
sub-regions of Sh2-252. Our analyses suggest that the region A is the youngest ($\sim$ 0.5 Myr), the regions B, C and E are of similar  
evolutionary stage ($\sim$ 1-2 Myr) and the small clusters NGC 2175s and Teu 136, located towards the east of Sh2-252  are slightly evolved 
($\sim$ 2-3 Myr). Morphology of the region in the {\it Spitzer} colour images as well as in the 1.1 mm map shows an almost semi-circular ring-like 
shape towards the western half of the complex. Indeed, we find a molecular shell composed of several clumps distributed around the main 
ionizing source (i.e., HD 42088), suggesting that the expansion of the \hii region is collecting the surrounding material, which gives rise to the 
semi-circular ring shape. We find several candidate YSOs distributed over the semi-circular molecular shell which is an evidence for the star 
formation activity within the shell.  Finally, by comparing the  age of the ionizing source, fragmentation time of the collected molecular 
shell and age of the YSOs, we suggest  collect and collapse scenario as one of the  the possible mechanisms responsible for the star formation within the 
shell. We observed the densest concentration of YSOs, (mostly Class I, $\sim$ 0.5 Myr) 
at the western outskirts of the complex, within a  molecular clump located between the C\hii regions A and B. The correlation between the molecular 
clump at this location with large number of Class I YSOs and the associated  water and methanol masers suggest that it is indeed  a site of 
cluster formation in a very early evolutionary stage, sandwiched between the two relatively evolved C\hii regions A and B. We conclude that the 
region is undergoing a complex star formation activity and there is a strong interplay between the radiation from the expanding \hii region to 
the surrounding molecular material. The region certainly deserves attention for high resolution molecular line observations to further explore its 
hidden structure.
% We also report the identification of a candidate runaway star of B1V spectral type located at the southern part of the complex. 

\section{Acknowledgments}

We are thankful to the anonymous referee 	for useful comments which  have improved the 	contents and presentation of the paper
significantly. This publication  makes use of  data from  the Two  
Micron All Sky Survey, which is  a joint project of  the University of  Massachusetts and the Infrared  
Processing  and   Analysis  Center/California  Institute  of Technology, funded by the National Aeronautics   
and Space Administration and the National  Science Foundation. This work is based on observations made with the
{\it Spitzer} Space Telescope, which is operated by the Jet Propulsion Laboratory, California Institute of Technology, 
under contract with NASA. This publication makes extensive use of data products from the HST archive at the Canadian
Astronomy Data Center (CADC), operated by the National Research Council of Canada with the support of the Canadian Space
Agency.  JJ is thankful for the financial support for part of this study through a stipend from  CSIR, India
and also the guidance  from the {\it Spitzer} support staff for the data analysis is highly acknowledged.

\begin{table*}
\caption{Physical parameters of YSOs from SED fits. The complete table is available in the electronic version. }
\label{sedpars}
\scriptsize
%\begin{tabular}{lllllllll}
\begin{tabular}{|p{.05in}p{.4in}p{.4in}p{0.68in}p{0.68in}p{0.68in}p{0.68in}p{0.9in}p{0.95in}|}
\hline
%\begin{deluxetable}{lccllllll}
%\tablecolumns{9}
%\tablewidth{0pt}
%\tablecaption{\bf \underline {Physical parameters of YSOs from SED fits}}
%\tablehead{

ID & $\alpha_{(2000)}$ & $\delta_{(2000)}$ & $M_*$      & $T_*$     & $t_*$     & $M_{disk}$  & $\dot{M}_{disk}$       & $\dot{M}_{env}$\\
   &  (deg)                 &  (deg)              &($M_{\odot}$)& ($10^4 K$)&($10^6$yr)&($M_{\odot}$) & ($10^{-7}$$M_{\odot}$/yr)& ($10^{-6}$$M_{\odot}$/yr)\\ 

%\startdata
\hline
1  &  92.242438  &   20.391663   &   2.1  $\pm$   0.5   &   0.77  $\pm$   0.29 &   4.42  $\pm$   3.28   &   0.026   $\pm$   0.028   &   3.542   $\pm$   2.896   &   0.002   $\pm$   0.003\\
2  &  92.066338  &   20.544426   &   1.4  $\pm$   1.0   &   0.43  $\pm$   0.12 &   0.06  $\pm$   0.23   &   0.019   $\pm$   0.040   &   5.907   $\pm$   5.553   &   126.0   $\pm$   116.0\\
3  &  92.196279  &   20.630317   &   4.5  $\pm$   0.3   &   0.75  $\pm$   0.10 &   0.74  $\pm$   0.23   &   0.001   $\pm$   0.001   &   0.038   $\pm$   0.037   &   0.400   $\pm$   0.562\\
4  &  92.464273  &   20.476278   &   3.1  $\pm$   0.5   &   1.17  $\pm$   0.15 &   4.85  $\pm$   1.77   &   0.003   $\pm$   0.007   &   0.028   $\pm$   0.028   &   0.000   $\pm$   0.000\\
5  &  92.068404  &   20.607574   &   3.7  $\pm$   0.6   &   1.34  $\pm$   0.14 &   4.00  $\pm$   2.17   &   0.011   $\pm$   0.022   &   1.256   $\pm$   1.186   &   1.127   $\pm$   17.51\\
\hline
\end{tabular}

\end{table*}

\bsp
%label{lastpage}


\begin{thebibliography}{}

\bibitem[\protect\citeauthoryear{{Aguirre}, {Ginsburg}, {Dunham}, {Drosback},
  {Bally}, {Battersby}, {Bradley}, {Cyganowski}, {Dowell}, {Evans} II, {Glenn},
  {Harvey}, {Rosolowsky}, {Stringfellow}, {Walawender} \& {Williams}}{{Aguirre}
  et~al.}{2011}]{aquirre2011}
{Aguirre} J.~E. et~al. 2011, \apjs,   192, 4

\bibitem[\protect\citeauthoryear{{Allen}, {Calvet}, {D'Alessio}, {Merin},
  {Hartmann}, {Megeath}, {Gutermuth}, {Muzerolle}, {Pipher}, {Myers} \&
  {Fazio}}{{Allen} et~al.}{2004}]{allen2004}
{Allen} L.~E. et~al.  2004, \apjs, 154, 363

\bibitem[\protect\citeauthoryear{{Allen}, {Pipher}, {Gutermuth}, {Megeath},
  {Adams}, {Herter}, {Williams}, {Goetz-Bixby}, {Allen} \& {Myers}}{{Allen}
  et~al.}{2008}]{allen2008}
{Allen} T.~S. et~al.  2008, \apj, 675, 491

\bibitem[\protect\citeauthoryear{{Bally}, {Testi}, {Sargent} \&
  {Carlstrom}}{{Bally} et~al.}{1998}]{bally1998}
{Bally} J.,  {Testi} L.,  {Sargent} A.,    {Carlstrom} J.,  1998, \aj, 116, 854

\bibitem[\protect\citeauthoryear{{Beerer}, {Koenig}, {Hora}, {Gutermuth},
  {Bontemps}, {Megeath}, {Schneider}, {Motte}, {Carey}, {Simon}, {Keto},
  {Smith}, {Allen}, {Fazio}, {Kraemer}, {Price}, {Mizuno}, {Adams},
  {Hern{\'a}ndez} \& {Lucas}}{{Beerer} et~al.}{2010}]{beerer2010}
{Beerer} I.~M. et~al. 2010, \apj, 720, 679

\bibitem[\protect\citeauthoryear{{Bisbas}, {W{\"u}nsch}, {Whitworth}, {Hubber}
  \& {Walch}}{{Bisbas} et~al.}{2011}]{bisbas2011}
{Bisbas} T.~G.,  {W{\"u}nsch} R.,  {Whitworth} A.~P.,  {Hubber} D.~A.,
  {Walch} S.,  2011, \apj, 736, 142

\bibitem[\protect\citeauthoryear{{Blaauw}}{{Blaauw}}{1961}]{blaauw1961}
{Blaauw} A.,  1961, Bull. Astron. Inst. Netherlands,  15, 265

\bibitem[\protect\citeauthoryear{{Bohlin}, {Savage} \& {Drake}}{{Bohlin}
  et~al.}{1978}]{bohlin1978}
{Bohlin} R.~C.,  {Savage} B.~D.,    {Drake} J.~F.,  1978, \apj, 224, 132

\bibitem[\protect\citeauthoryear{{Chavarr{\'{\i}}a}, {Allen}, {Hora}, {Brunt}
  \& {Fazio}}{{Chavarr{\'{\i}}a} et~al.}{2008}]{chavarria2008}
{Chavarr{\'{\i}}a} L.~A.,  {Allen} L.~E.,  {Hora} J.~L.,  {Brunt} C.~M.,
  {Fazio} G.~G.,  2008, \apj, 682, 445

\bibitem[\protect\citeauthoryear{{Chauhan}, {Pandey}, {Ogura}, {Ojha}, {Bhatt},
  {Ghosh} \& {Rawat}}{{Chauhan} et~al.}{2009}]{chauhan2009}
{Chauhan} N.,  {Pandey} A.~K.,  {Ogura} K.,  {Ojha} D.~K.,  {Bhatt} B.~C.,
  {Ghosh} S.~K.,    {Rawat} P.~S.,  2009, \mnras, 396, 964

\bibitem[\protect\citeauthoryear{{Cohen}, {Staveley-Smith} \& {Green}}{{Cohen}
  et~al.}{2003}]{cohen2003}
{Cohen} M.,  {Staveley-Smith} L.,    {Green} A.,  2003, \mnras, 340, 275

\bibitem[\protect\citeauthoryear{{Conti}, {Leep} \& {Lorre}}{{Conti}
  et~al.}{1977}]{conti1977}
{Conti} P.~S.,  {Leep} E.~M.,    {Lorre} J.~J.,  1977, \apj, 214, 759

\bibitem[\protect\citeauthoryear{{Cutri}, {Skrutskie}, {van Dyk}, {Beichman},
  {Carpenter}, {Chester}, {Cambresy}, {Evans}, {Fowler}, {Gizis}, {Howard},
  {Huchra}, {Jarrett}, {Kopan}, {Kirkpatrick}, {Light}, {Marsh}, {McCallon},
  {Schneider}, {Stiening}, {Sykes}}{{Cutri} et~al.}{2003}]{cutri2003}
{Cutri} R.~M. et~al. 2003, {2MASS All Sky Catalog of   point sources,
 NASA/IPAC Infrared Science Archive, http://irsa.ipac.caltech.edu/applications/Gator/}

\bibitem[\protect\citeauthoryear{{Deharveng}, {Lefloch}, {Kurtz}, {Nadeau},
  {Pomar{\`e}s}, {Caplan} \& {Zavagno}}{{Deharveng}
  et~al.}{2008}]{deharveng2008}
{Deharveng} L.,  {Lefloch} B.,  {Kurtz} S.,  {Nadeau} D.,  {Pomar{\`e}s} M.,
  {Caplan} J.,    {Zavagno} A.,  2008, \aap, 482, 585

\bibitem[\protect\citeauthoryear{{Deharveng}, {Zavagno}, {Schuller}, {Caplan},
  {Pomar{\`e}s} \& {De Breuck}}{{Deharveng} et~al.}{2009}]{deharveng2009}
{Deharveng} L.,  {Zavagno} A.,  {Schuller} F.,  {Caplan} J.,  {Pomar{\`e}s} M.,
     {De Breuck} C.,  2009, \aap, 496, 177

\bibitem[\protect\citeauthoryear{{Evans} II, {Dunham}, {J{\o}rgensen}, {Enoch},
  {Mer{\'{\i}}n}, {van Dishoeck}, {Alcal{\'a}}, {Myers}, {Stapelfeldt},
  {Huard}, {Allen}, {Harvey}, {van Kempen}, {Blake}, {Koerner}, {Mundy},
  {Padgett} \& {Sargent}}{{Evans} et~al.}{2009}]{evans2009}
{Evans} II N.~J. et~al.  2009, \apjs, 181, 321

\bibitem[\protect\citeauthoryear{{Fang}, {van Boekel}, {Wang}, {Carmona},
  {Sicilia-Aguilar} \& {Henning}}{{Fang} et~al.}{2009}]{fang2009}
{Fang} M.,  {van Boekel} R.,  {Wang} W.,  {Carmona} A.,  {Sicilia-Aguilar} A.,
    {Henning} T.,  2009, \aap, 504, 461

\bibitem[\protect\citeauthoryear{{Felli}, {Habing} \& {Israel}}{{Felli}
  et~al.}{1977}]{felli1977}
{Felli} M.,  {Habing} H.~J.,    {Israel} F.~P.,  1977, \aap, 59, 43

\bibitem[\protect\citeauthoryear{{Flaherty}, {Pipher}, {Megeath}, {Winston},
  {Gutermuth}, {Muzerolle}, {Allen} \& {Fazio}}{{Flaherty}
  et~al.}{2007}]{flaherty2007}
{Flaherty} K.~M.,  {Pipher} J.~L.,  {Megeath} S.~T.,  {Winston} E.~M.,
  {Gutermuth} R.~A.,  {Muzerolle} J.,  {Allen} L.~E.,    {Fazio} G.~G.,  2007,
  \apj, 663, 1069

\bibitem[\protect\citeauthoryear{{Fountain}, {Gary} \& {Odell}}{{Fountain}
  et~al.}{1983}]{fountain1983}
{Fountain} W.~F.,  {Gary} G.~A.,    {Odell} C.~R.,  1983, \apj, 273, 639

\bibitem[\protect\citeauthoryear{{Gies} \& {Bolton}}{{Gies} \&
  {Bolton}}{1986}]{gies1986}
{Gies} D.~R.,  {Bolton} C.~T.,  1986, \apjs, 61, 419

\bibitem[\protect\citeauthoryear{{Gutermuth}, {Megeath}, {Myers}, {Allen},
  {Pipher} \& {Fazio}}{{Gutermuth} et~al.}{2009}]{gutermuth2009}
{Gutermuth} R.~A.,  {Megeath} S.~T.,  {Myers} P.~C.,  {Allen} L.~E.,  {Pipher}
  J.~L.,    {Fazio} G.~G.,  2009, \apjs, 184, 18

\bibitem[\protect\citeauthoryear{{Gutermuth}, {Myers}, {Megeath}, {Allen},
  {Pipher}, {Muzerolle}, {Porras}, {Winston} \& {Fazio}}{{Gutermuth}
  et~al.}{2008}]{gutermuth2008}
{Gutermuth} R.~A. et~al.,  2008,   \apj, 674, 336

\bibitem[\protect\citeauthoryear{{Harvey}, {Mer{\'{\i}}n}, {Huard}, {Rebull},
  {Chapman}, {Evans} II \& {Myers}}{{Harvey} et~al.}{2007}]{harvey2007}
{Harvey} P.,  {Mer{\'{\i}}n} B.,  {Huard} T.~L.,  {Rebull} L.~M.,  {Chapman}
  N.,  {Evans} II N.~J.,    {Myers} P.~C.,  2007, \apj, 663, 1149

\bibitem[\protect\citeauthoryear{{Harvey}, {Chapman}, {Lai}, {Evans} II,
  {Allen}, {J{\o}rgensen}, {Mundy}, {Huard}, {Porras}, {Cieza}, {Myers},
  {Mer{\'{\i}}n}, {van Dishoeck}, {Young}, {Spiesman}, {Blake}, {Koerner},
  {Padgett}, {Sargent} \& {Stapelfeldt}}{{Harvey} et~al.}{2006}]{harvey2006}
{Harvey} P.~M. et~al. 2006, \apj, 644, 307

\bibitem[\protect\citeauthoryear{{Hatchell}, {Fuller}, {Richer}, {Harries} \&
  {Ladd}}{{Hatchell} et~al.}{2007}]{hatchell2007}
{Hatchell} J.,  {Fuller} G.~A.,  {Richer} J.~S.,  {Harries} T.~J.,    {Ladd}
  E.~F.,  2007, \aap, 468, 1009

\bibitem[\protect\citeauthoryear{{Jacoby}, {Hunter} \& {Christian}}{{Jacoby}
  et~al.}{1984}]{jacoby1984}
{Jacoby} G.~H.,  {Hunter} D.~A.,    {Christian} C.~A.,  1984, \apjs, 56, 257

\bibitem[\protect\citeauthoryear{{Jose}, {Pandey}, {Ogura}, {Ojha}, {Bhatt},
  {Samal}, {Chauhan}, {Sahu} \& {Rawat}}{{Jose} et~al.}{2011}]{jose2011}
{Jose} J. et~al.  2011, \mnras,
  411, 2530

\bibitem[\protect\citeauthoryear{{Jose}, {Pandey}, {Ogura}, {Samal}, {Ojha},
  {Bhatt}, {Chauhan}, {Eswaraiah}, {Mito}, {Kobayashi} \& {Yadav}}{{Jose}
  et~al.}{2012}]{jose2012}
{Jose} J. et~al. 2012, \mnras, 424, 2486

\bibitem[\protect\citeauthoryear{{Jose}, {Pandey}, {Ojha}, {Ogura}, {Chen},
  {Bhatt}, {Ghosh}, {Mito}, {Maheswar} \& {Sharma}}{{Jose}
  et~al.}{2008}]{jose2008}
{Jose} J. et~al.  2008,  \mnras, 384, 1675

\bibitem[\protect\citeauthoryear{{Koempe}, {Baudry}, {Joncas} \&
  {Wouterloot}}{{Koempe} et~al.}{1989}]{kompe1989}
{Koempe} C.,  {Baudry} A.,  {Joncas} G.,    {Wouterloot} J.~G.~A.,  1989, \aap,
  221, 295

\bibitem[\protect\citeauthoryear{{Lada}, {Blitz}, {Reid} \& {Moran}}{{Lada}
  et~al.}{1981}]{lada1981}
{Lada} C.~J.,  {Blitz} L.,  {Reid} M.~J.,    {Moran} J.~M.,  1981, \apj, 243,
  769

\bibitem[\protect\citeauthoryear{{Lada} \& {Wilking}}{{Lada} \&
  {Wilking}}{1984}]{lada1984}
{Lada} C.~J.,  {Wilking} B.~A.,  1984, \apj, 287, 610

\bibitem[\protect\citeauthoryear{{Lada} \& {Wooden}}{{Lada} \&
  {Wooden}}{1979}]{lada1979}
{Lada} C.~J.,  {Wooden} D.,  1979, \apj, 232, 158

\bibitem[\protect\citeauthoryear{{Lefloch}, {Lazareff} \& {Castets}}{{Lefloch}
  et~al.}{1997}]{lefloch1997}
{Lefloch} B.,  {Lazareff} B.,    {Castets} A.,  1997, \aap, 324, 249

\bibitem[\protect\citeauthoryear{{Mdzinarishvili} \&
  {Chargeishvili}}{{Mdzinarishvili} \&
  {Chargeishvili}}{2005}]{mdzinarishvili2005}
{Mdzinarishvili} T.~G.,  {Chargeishvili} K.~B.,  2005, \aap, 431, L1

\bibitem[\protect\citeauthoryear{{Megeath}, {Allen}, {Gutermuth}, {Pipher},
  {Myers}, {Calvet}, {Hartmann}, {Muzerolle} \& {Fazio}}{{Megeath}
  et~al.}{2004}]{megeath2004}
{Megeath} S.~T. et~al.,  2004, \apjs, 154, 367

\bibitem[\protect\citeauthoryear{{Ojha}, {Samal}, {Pandey}, {Bhatt}, {Ghosh},
  {Sharma}, {Tamura}, {Mohan} \& {Zinchenko}}{{Ojha} et~al.}{2011}]{ojha2011}
{Ojha} D.~K. et~al.,  2011, \apj, 738,   156

\bibitem[\protect\citeauthoryear{{Pandey}, {Sharma}, {Ogura}, {Ojha}, {Chen},
  {Bhatt} \& {Ghosh}}{{Pandey} et~al.}{2008}]{pandey2008}
{Pandey} A.~K.,  {Sharma} S.,  {Ogura} K.,  {Ojha} D.~K.,  {Chen} W.~P.,
  {Bhatt} B.~C.,    {Ghosh} S.~K.,  2008, \mnras, 383, 1241

\bibitem[\protect\citeauthoryear{{Pomar{\`e}s}, {Zavagno}, {Deharveng},
  {Cunningham}, {Jones}, {Kurtz}, {Russeil}, {Caplan} \&
  {Comer{\'o}n}}{{Pomar{\`e}s} et~al.}{2009}]{pomares2009}
{Pomar{\`e}s} M.,  et~al.,  2009,   \aap, 494, 987

\bibitem[\protect\citeauthoryear{{Robitaille}, {Whitney}, {Indebetouw} \&
  {Wood}}{{Robitaille} et~al.}{2007}]{robitaille2007}
{Robitaille} T.~P.,  {Whitney} B.~A.,  {Indebetouw} R.,    {Wood} K.,  2007,
  \apjs, 169, 328

\bibitem[\protect\citeauthoryear{{Robitaille}, {Whitney}, {Indebetouw}, {Wood}
  \& {Denzmore}}{{Robitaille} et~al.}{2006}]{robitaille2006}
{Robitaille} T.~P.,  {Whitney} B.~A.,  {Indebetouw} R.,  {Wood} K.,
  {Denzmore} P.,  2006, \apjs, 167, 256

Schmidt-Kaler Th., 1982, in Schaifers K., Voigt H. H., Landolt H., eds,
   Landolt-Bornstein, Vol. 2b, Springer, Berlin, p. 19

\bibitem[\protect\citeauthoryear{{Sharpless}}{{Sharpless}}{1959}]{sharpless195%
9}
{Sharpless} S.,  1959, \apjs, 4, 257

\bibitem[\protect\citeauthoryear{{Siess}, {Dufour} \& {Forestini}}{{Siess}
  et~al.}{2000}]{siess2000}
{Siess} L.,  {Dufour} E.,    {Forestini} M.,  2000, \aap, 358, 593

\bibitem[\protect\citeauthoryear{{Smith}, {Povich}, {Whitney}, {Churchwell},
  {Babler}, {Meade}, {Bally}, {Gehrz}, {Robitaille} \& {Stassun}}{{Smith}
  et~al.}{2010}]{smith2010}
{Smith} N. et~al., 2010, \mnras, 406, 952

\bibitem[\protect\citeauthoryear{{Stone}}{{Stone}}{1982}]{stone1982}
{Stone} R.~C.,  1982, \apj, 261, 208

\bibitem[\protect\citeauthoryear{{Szymczak}, {Kus} \& {Hrynek}}{{Szymczak}
  et~al.}{2000}]{szymczak2000}
{Szymczak} M.,  {Kus} A.~J.,    {Hrynek} G.,  2000, \mnras, 312, 211

\bibitem[\protect\citeauthoryear{{Tej}, {Ojha}, {Ghosh}, {Kulkarni}, {Verma},
  {Vig} \& {Prabhu}}{{Tej} et~al.}{2006}]{tej2006}
{Tej} A.,  {Ojha} D.~K.,  {Ghosh} S.~K.,  {Kulkarni} V.~K.,  {Verma} R.~P.,
  {Vig} S.,    {Prabhu} T.~P.,  2006, \aap, 452, 203

\bibitem[\protect\citeauthoryear{{Vacca}, {Garmany} \& {Shull}}{{Vacca}
  et~al.}{1996}]{vacca1996}
{Vacca} W.~D.,  {Garmany} C.~D.,    {Shull} J.~M.,  1996, \apj, 460, 914

\bibitem[\protect\citeauthoryear{{Walborn} \& {Fitzpatrick}}{{Walborn} \&
  {Fitzpatrick}}{1990}]{walborn1990}
{Walborn} N.~R.,  {Fitzpatrick} E.~L.,  1990, \pasp, 102, 379

\bibitem[\protect\citeauthoryear{{Whitney}, {Indebetouw}, {Bjorkman} \&
  {Wood}}{{Whitney} et~al.}{2004}]{whitney2004}
{Whitney} B.~A.,  {Indebetouw} R.,  {Bjorkman} J.~E.,    {Wood} K.,  2004,
  \apj, 617, 1177

\bibitem[\protect\citeauthoryear{{Whitney}, {Wood}, {Bjorkman} \&
  {Cohen}}{{Whitney} et~al.}{2003b}]{whitney2003b}
{Whitney} B.~A.,  {Wood} K.,  {Bjorkman} J.~E.,    {Cohen} M.,  2003, \apj,
  598, 1079

\bibitem[\protect\citeauthoryear{{Whitney}, {Wood}, {Bjorkman} \&
  {Wolff}}{{Whitney} et~al.}{2003a}]{whitney2003a}
{Whitney} B.~A.,  {Wood} K.,  {Bjorkman} J.~E.,    {Wolff} M.~J.,  2003, \apj,
  591, 1049

\bibitem[\protect\citeauthoryear{{Whitworth}, {Bhattal}, {Chapman}, {Disney} \&
  {Turner}}{{Whitworth} et~al.}{1994}]{whitworth1994}
{Whitworth} A.~P.,  {Bhattal} A.~S.,  {Chapman} S.~J.,  {Disney} M.~J.,
  {Turner} J.~A.,  1994, \mnras, 268, 291

\bibitem[\protect\citeauthoryear{{Yamaguchi}, {Mizuno}, {Onishi}, {Mizuno} \&
  {Fukui}}{{Yamaguchi} et~al.}{2001a}]{yamaguchi2001a}
{Yamaguchi} R.,  {Mizuno} N.,  {Onishi} T.,  {Mizuno} A.,    {Fukui} Y.,
  2001a, \apjl, 553, L185

\bibitem[\protect\citeauthoryear{{Yamaguchi}, {Mizuno}, {Onishi}, {Mizuno} \&
  {Fukui}}{{Yamaguchi} et~al.}{2001b}]{yamaguchi2001b}
{Yamaguchi} R.,  {Mizuno} N.,  {Onishi} T.,  {Mizuno} A.,    {Fukui} Y.,
  2001b, \pasj, 53, 959

\end{thebibliography}
\end{document}